\documentclass[draft]{agujournal2019}
\usepackage[T1]{fontenc}
\usepackage{url}
\usepackage{lineno}
\usepackage[inline]{trackchanges}
\usepackage{soul}
\draftfalse

\usepackage{amsmath}
\usepackage{amsfonts}
\usepackage{amssymb}
\usepackage{graphicx}
\usepackage{booktabs}
\usepackage{tabularx}
\usepackage{xcolor}

\graphicspath{{./figures/}}

\journalname{Space Weather}

\begin{document}

\title{Aurora Hunter: A Two-Stage Framework for Probabilistic Visibility Forecasting}

\authors{Z. Ge\affil{1,2,3},
  C. Zhang\affil{1},
  H. Li\affil{1},
  H. Zhang\affil{1},
  W. Zhou\affil{4},
  W. Gu\affil{1},
  Z.-M. Wang\affil{1,5,6}}

\affiliation{1}{College of Physics and Optoelectronic Engineering,
  Ocean University of China, Qingdao 266100, China}
\affiliation{2}{Shanghai Qi Zhi Institute, Shanghai 200232, China}
\affiliation{3}{Department of Physics, National University of Singapore,
  Singapore 117551, Singapore}
\affiliation{4}{School of Mathematics and Computer Science,
  Yunnan Minzu University, Kunming 650504, People's Republic of China}
\affiliation{5}{Engineering Research Center of Advanced Marine Physical
  Instruments and Equipment, Ministry of Education, Qingdao 266100, China}
\affiliation{6}{Qingdao Key Laboratory of Optics and Optoelectronics,
  Qingdao 266100, China}

\correspondingauthor{Zhao-Ming Wang}{wangzhaoming@ouc.edu.cn}

\begin{keypoints}
\item A two-stage cascade decouples aurora occurrence from observing
  conditions, matching a single-stage model with each factor
  interpretable.
\item The cascade achieves ROC-AUC 0.958 at Troms\o\ and 0.933 at
  independent Kiruna, improving on the occurrence stage alone by 0.095 and
  0.097.
\item Pairing stations in oval-relative coordinates constrains the
  occurrence profile and yields hemispheric maps with a measured validity
  domain.
\end{keypoints}

\begin{abstract}
Aurora visibility at a given location requires two physically distinct
conditions to hold at once: aurora occurring overhead, governed by solar
wind--magnetosphere coupling, and observing conditions that permit detection,
governed by cloud cover and moonlight.  Approaches that conflate the two weaken
the space-weather signal and limit cross-site generalizability.

We develop Aurora Hunter, a two-stage cascade that separates
occurrence prediction from observing-condition
assessment.  Stage~1 uses 43 physics-driven features to
predict $P(\text{aurora identifiable in all-sky
images})$ via gradient-boosted trees trained on joint
Troms\o{}+Kiruna data ($\sim$16{,}600 hours, 2015--2023).  Stage~2 models
$P(\text{unobscured image class} \mid \text{identifiable})$
with logistic regression on 15 observing-condition
features, trained on hours with identifiable aurora.  The cascade
$P(\text{visible}) = P(\text{identifiable}) \times P(\text{unobscured} \mid \text{identifiable})$
achieves retrospective area under the receiver operating characteristic curve (ROC-AUC) of 0.958 (Troms\o\ test,
2019--2020) and 0.933 (independent Kiruna, 2024), improving on the occurrence
stage used alone by $+0.095$ and $+0.097$.  Transfer to Skibotn, the one
station withheld entirely (2022--2025), is limited.  SHAP
analysis identifies magnetic local time, the Kp$\times$nightside interaction and
the three-hour mean Kp as the dominant features (44\% of attribution),
consistent with auroral oval physics.  Hemisphere-wide occurrence maps
combine the Stage~1 amplitude, on a clear-sky basis, with the Feldstein oval
parameterization.  By providing location-specific
visibility probabilities with measured reliability rather than
coarse geomagnetic indices, the framework links space weather research to
practical observation planning.  An operational
proof of concept is available at \url{https://aurora-hunter.onrender.com}.
\end{abstract}

\section*{Plain Language Summary}
Seeing the northern lights requires two things at the same time: active
space weather sending energetic particles into the atmosphere, and clear
skies overhead.  We built a system called Aurora Hunter that estimates
both conditions separately and then combines them into a single probability
of getting a clear view of the aurora.  The space weather part
uses 43 features derived from solar wind measurements to predict whether
aurora is overhead at all.  The sky-conditions part uses
cloud cover and moon brightness to predict whether the aurora would
show up clearly.  Multiplying the two probabilities together gives one
number for a particular place and a particular hour, whereas the usual rule of thumb
pairs a geomagnetic activity level with a separate glance at the
weather.

We train and test the system against what an automatic classifier found in
all-sky camera photographs at three sites in northern Scandinavia.  Because each
site enters the model by its position relative to the auroral oval rather than
by its name, the same occurrence estimate extends to hour-by-hour maps of the
whole Northern Hemisphere, and the paper states the range over which those maps
have been checked.

\section{Introduction}
\label{sec:intro}

The aurora borealis is a prominent manifestation of solar
wind--magnetosphere--ionosphere coupling
\cite{akasofu1964auroral}, and its prediction has attracted growing
attention from both the space weather community and the aurora tourism
industry \cite{case2015aurora}.  The practical forecasting goal differs
between the two communities: for space weather research it is whether aurora is
physically occurring, governed by solar wind--magnetosphere coupling, while for
aurora tourism in high-latitude regions it is
\emph{naked-eye visibility}: whether the aurora will be
visible to an observer at a given location on a given night.  To meet
this demand, a distinctive professional practice has developed in
aurora tourism regions: local guides colloquially known as \emph{aurora
hunters} drive groups of tourists across the terrain in real-time
pursuit of auroral displays, adjusting their routes as conditions
evolve through the night.  In practice, both professional guides and
the consumer forecast applications that serve the broader public base
their decisions on two readily observable indicators: the Kp
geomagnetic index \cite{bartels1939kp} as a measure of magnetospheric
disturbance level, and cloud-cover maps to identify sky-transparency
windows.  This two-factor heuristic, Kp plus cloud cover, represents
the current practical state of the art for tourist aurora prediction.

This heuristic has proven useful in practice, but its inherent
limitations constrain forecast accuracy.  Kp alone does not encode
the position of the auroral oval relative to the observer's magnetic
latitude, and cloud cover assessment typically does not account for
lunar illumination, which can suppress detection even under otherwise
clear skies.
The fundamental forecasting challenge lies in the intertwining of two
physically distinct questions: (i)~whether aurora is physically
occurring above the site, and (ii)~whether local observing conditions
permit detection.  To date, most practical and scientific
forecasting approaches have not explicitly separated these two
conditions.

\subsection{The Two-Condition Problem}

Aurora visibility requires the simultaneous satisfaction of two
physically distinct conditions.  The first is \emph{aurora occurrence}:
aurora must be physically present overhead, which depends on
solar wind--magnetosphere coupling, geomagnetic activity (quantified
by the Kp index), and the observer's position relative to the auroral
oval, all driven by space weather on timescales of minutes to hours.
The second is \emph{clear observing conditions}: the aurora must be
detectable from the ground, which requires sufficiently transparent
skies (absence of cloud cover at all altitudes) together with
sufficiently low lunar illumination to distinguish faint auroral
structures from sky background.  This second condition is driven by
tropospheric meteorology and the lunar cycle, operating on timescales
largely independent of space weather.

The probability that an hour yields unobscured auroral imagery is therefore
the \emph{product} of these two probabilities:
\begin{linenomath*}
\begin{equation}
  P(\text{visible}) = P(\text{identifiable}) \times
    P(\text{unobscured} \mid \text{identifiable}).
  \label{eq:cascade}
\end{equation}
\end{linenomath*}
Both factors are defined from all-sky image classifications in
Section~\ref{sec:labels}, where the sense in which we use \emph{visibility} is
made precise.

\begin{figure*}[t!]
\noindent\includegraphics[width=\columnwidth]{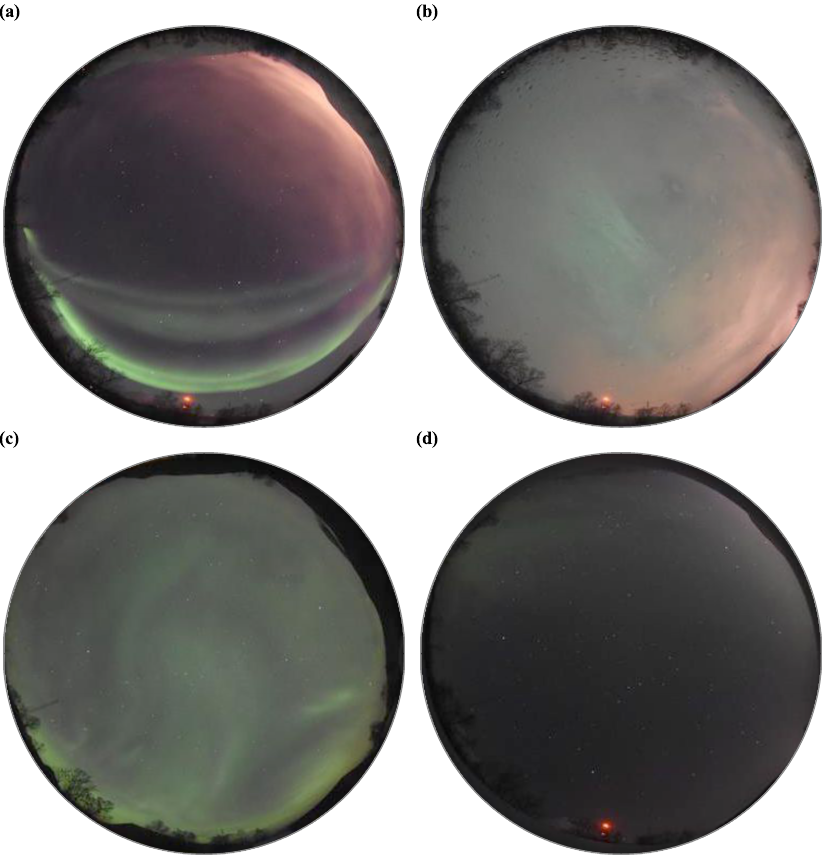}
\caption{Representative all-sky camera images from the Troms\o\ AI
  dataset \cite{nanjo2022automated}, four of its eight classes:
  (a)~Arc, an auroral arc under a dark, largely cloud-free sky;
  (b)~AC, aurora identified through cloud;
  (c)~Diffuse, diffuse aurora under a dark sky;
  (d)~Clear, a dark sky with no aurora.
  Stage~1 separates (a,\,b,\,c) from (d); Stage~2 separates
  (a,\,c) from (b).}
\label{fig:conditions_panel}
\end{figure*}

Figure~\ref{fig:conditions_panel} illustrates this distinction with
representative images from the Troms\o\ AI training dataset: (a)~an
auroral arc (Arc), (b)~aurora through cloud (AC),
(c)~diffuse aurora (Diffuse), and (d)~a dark sky with no aurora
(Clear).  Panels (a)--(c)
all show aurora identified in the image (in (b) through cloud); (a) and (c)
differ in auroral morphology rather than in sky condition, and panel (d) contains
no aurora.  These
four classes motivate our two-stage framework, which we
formalize next and then compare to existing approaches.

\subsection{Related Work and Motivation}

Existing aurora prediction approaches have been developed along three
main lines, each addressing different aspects of the forecasting
problem but not yet fully capturing the two-condition structure
described above.  First, Kp-based rule methods
combine a Kp threshold with local cloud cover forecasts
\cite{bartels1939kp,carbary2005kp} and form the explicit basis of the
professional aurora hunter heuristic.  While intuitive, Kp provides
only 3-hour temporal resolution and cannot resolve the fine-scale
substorm dynamics that determine whether aurora appears at a specific
site during a specific hour \cite{akasofu1964auroral,rostoker1980substorm}.
Furthermore, neither a Kp threshold nor cloud cover encodes the
observer's magnetic latitude relative to the oval, or the effect of
lunar illumination on detection.  \citeA{feng2025kp}
predicted UV auroral oval boundaries from Kp, and \citeA{wang2023kp}
applied XGBoost with solar imagery for 3-day Kp
forecasting (F1~$\approx 0.96$), but these systems predict geomagnetic
indices or oval boundaries, not site-specific probabilities of
unobscured auroral imagery.

Second, physics-based models such as OVATION Prime
\cite{newell2014ovation} predict global auroral precipitation flux
from the Newell coupling function \cite{newell2007}.  These models
are physically grounded and have demonstrated skill for oval location
(ROC-AUC~$\approx 0.82$) when validated against satellite data
\cite{mooney2021evaluating} and ground-based visible-aurora
observations \cite{machol2012auroral}.  However, these validation
metrics do not translate directly to ground-level aurora detection,
and OVATION entirely omits atmospheric observing conditions.

Third, end-to-end machine-learning approaches use deep networks to map
space-weather inputs to auroral morphology or occurrence.  \citeA{hu2021modeling}
used GRNN/CGAN to predict Polar UVI images
from interplanetary parameters, while \citeA{han2020prediction}
applied a deep learning model to predict
auroral oval boundaries from solar wind parameters.  Aurora image
classification has been advanced by convolutional neural networks
including systems trained on THEMIS all-sky data
\cite{clausen2018thermis,sado2022tame,johnson2024themis} and on Chinese
polar all-sky imagery \cite{zhong2018aurora}.  \citeA{nanjo2022automated}
developed a ResNet-50 real-time aurora
classifier (F1~$= 93.4\%$) that produces fine-grained per-image
classifications including the Aurora but Cloudy (AC)
category, the same system we use for ground-truth labels.  However,
when such classifiers are used to define a single binary visibility
target, downstream forecasting systems may inadvertently mix occurrence
with visibility: a negative training sample could represent either
absent aurora or aurora obscured by clouds, an ambiguity
that can weaken the learned physical relationships and limit cross-site
transferability.

\subsection{Our Approach}

We address these limitations with Aurora Hunter, a two-stage cascade
that explicitly separates the two physical conditions.  The key
contribution is the principled decomposition of the prediction problem:
Stage~1 learns $P(\text{identifiable})$ from physics-driven
features (space weather inputs, magnetic coordinates, and temporal lag
structures) with training labels that treat AC hours
as positive examples of aurora occurrence, thereby decoupling the
physics signal from weather.  Because Stage~1 relies only on space
weather and magnetic-coordinate inputs, it can be applied at any
geographic location without local ground-based observations.  Stage~2
then learns the conditional probability
$P(\text{unobscured} \mid \text{identifiable})$ from cloud-cover
and lunar-illumination features, trained exclusively on hours when
aurora is known to be occurring.  This isolation ensures that each
stage learns from the physically appropriate signal.

Multi-station training on joint Troms\o{}+Kiruna data (Section~\ref{sec:methods})
supports application at sites that contribute no training data; we quantify that
transfer at Skibotn, the one station withheld from training entirely, whose 2022--2025
record spans the Solar Cycle~25 maximum.  Ground truth is
provided by the Troms\o\ AI per-hour image classifications
\cite{nanjo2022automated} described in Section~\ref{sec:data_aurora}.
The framework is named \emph{Aurora Hunter} in reference to the
professional aurora guides whose two-factor assessment (geomagnetic
activity and sky conditions) it formalizes into spatially resolved
probability maps.

Aurora Hunter has been deployed as a real-time web application
(\url{https://aurora-hunter.onrender.com}), providing hourly updated
visibility maps over the magnetic latitude band spanned by the training
stations, with locations outside that band flagged as extrapolation.  The present paper
documents the framework, datasets and training protocol, and evaluates them
retrospectively on withheld data at three Nordic sites.
The deployed prototype ingests near-real-time data streams that differ from the
archival datasets studied here, but it runs the same models by the same logic
(\ref{app:operational}).

\section{Observational Data}
\label{sec:data}

\subsection{Ground-Truth Aurora Classifications}
\label{sec:data_aurora}

Our primary source of aurora ground truth is the Troms\o\ AI all-sky
camera classification system \cite{nanjo2022automated}, which applies
a ResNet-50 convolutional neural network to \emph{individual} all-sky
images recorded through a circular fish-eye lens at Ramfjordmoen Research
Station, Troms\o, Norway.  For each image the classifier returns a
probability distribution over eight classes: aurora arc (Arc),
discrete aurora (Discrete), diffuse aurora (Diffuse),
Aurora but Cloudy (AC), Aurora but Bright (AB), clear
sky (Clear), cloudy sky (Cloudy), and Dusk \& Dawn
(DD).  The
Arc, Discrete, Diffuse, AC and
AB classes enter our label construction
(Section~\ref{sec:labels}).

Over the years used here, the Troms\o\ images were acquired every
30~s with an 8~s exposure \cite{nanjo2022automated}, giving up to 120
images per hour.  We form one record per clock hour by averaging the
per-image class probabilities over the images available in that hour, so
that the eight resulting hourly values are expressed in percent and sum
to 100.  Each record retains the number of contributing images, allowing
hours with sparse coverage to be identified.

The AC and AB categories are particularly significant
for our framework.  In both, aurora is identified in the image:
AC denotes aurora identified through cloud, and AB
aurora identified against a background brightened by moonlight or
sunlight \cite{nanjo2022automated}.  They differ from Arc,
Discrete and Diffuse in the state of the sky under
which the aurora was recorded.  Other all-sky classification schemes
carry no equivalent label: the six-class Oslo Auroral THEMIS (OATH)
scheme treats cloud and moon as non-auroral classes
\cite{clausen2018thermis,sado2022tame}, and the four-class scheme of
\citeA{zhong2018aurora} covers auroral morphology only.
\citeA{nanjo2022automated} define AC and AB but
exclude them from the observable classes used for their published
occurrence rate.  We instead retain them as positive Stage~1 samples, so
that hours in which aurora was detected are not counted as hours without
aurora.

The three primary observation sites used in this study are listed below.
Magnetic coordinates are given in the Altitude-Adjusted Corrected
Geomagnetic (AACGM-v2) system at 110~km as magnetic latitude (MLAT) and magnetic
local time (MLT), computed as described in Section~\ref{sec:data_coords}; the MLAT
quoted for each site is the mean over that site's own record, within which secular
variation moves it by less than $0.2^{\circ}$.  Times are universal time
(UT).

\begin{enumerate}
  \item Troms\o\ (Ramfjordmoen), Norway (geographic:
        69.6$^{\circ}$N, 19.2$^{\circ}$E; AACGM magnetic:
        MLAT~$\approx 66.96^{\circ}$, MLT
        offset~$\approx +1.9$~h from UT).  Data span 2015--2020, providing
        17,698 aligned hourly samples, of which 17,648 carry complete inputs and
        enter the splits of Table~\ref{tab:data_stats}.  Used for model
        training (2015--2017), validation (2018), and testing
        (2019--2020).

  \item Kiruna, Sweden (geographic: 67.84$^{\circ}$N,
        20.41$^{\circ}$E; MLAT~$\approx 65.24^{\circ}$).  Data span 2020--2024, providing
        8,885 hourly samples.  Used for independent geographic
        validation.

  \item Skibotn, Norway (geographic: 69.38$^{\circ}$N,
        20.36$^{\circ}$E; MLAT~$\approx 66.87^{\circ}$).  Data span 2022--2025, providing
        6,577 aligned hourly samples, 6,444 with complete inputs.  Withheld entirely as an independent test
        site covering the Solar Cycle~25 ascending and maximum phases,
        which are absent from the training data.
\end{enumerate}

The three stations span a narrow magnetic latitude range
(MLAT 65.24$^{\circ}$--66.96$^{\circ}$) but sample different phases of the
solar cycle, with the Troms\o\ record separated from the other two by 2--7
years.

All three stations are part of the same Troms\o\ AI system: color all-sky
imagers whose individual frames are classified by the same ResNet-50 network
into the same eight classes, so the target is defined identically at every
site.  Acquisition cadence differs (the median is 120 frames per hour at
Troms\o\ and Skibotn against 60 at Kiruna), which affects only the number of
frames averaged into each hourly class fraction, and each record retains its
frame count.  Because the classifier was trained on Troms\o\ images, its
per-class behavior can shift at the other sites; Section~\ref{sec:significance} and
Figure~\ref{fig:differential_case} compare the two sites' image classes hour by
hour, and the transfer results of
Table~\ref{tab:three_station} should be read with that caveat.

Two further datasets are introduced in Section~\ref{sec:global_method} and take
no part in training or in the cascade evaluation: human-labeled all-sky frames from
Ny-\r{A}lesund at MLAT $76.8^{\circ}$, and calibrated 557.7~nm intensities from an
independent photometric network.  They are used only to test the latitude profile of
the hemispheric occurrence field, nine degrees poleward of the stations above.

\subsection{Space Weather Parameters}
\label{sec:data_sw}

Solar wind and geomagnetic parameters are obtained from the NASA OMNI
1-hour dataset \cite{king2005omni} at hourly resolution.  We use OMNI
time in UT and align each station-hour to the corresponding OMNI
1-hour bin by UT.  We use: interplanetary magnetic field (IMF)
components $B_x$, $B_y$, $B_z$ in geocentric solar magnetospheric
(GSM) coordinates (nT); solar wind flow
speed $V_{sw}$ (km~s$^{-1}$) and proton density $n_p$ (cm$^{-3}$);
geomagnetic indices Kp and Dst (nT); and the derived Newell coupling
function $dN/dt$ \cite{newell2007}, empirical hemispheric power input
(HPI), and dynamic pressure $P_{\rm dyn}$ defined as
\begin{linenomath*}
\begin{equation}
  P_{\rm dyn} = \rho V_{sw}^2, \qquad \rho = m_p n_p,
  \label{eq:pdyn}
\end{equation}
\end{linenomath*}
where $\rho$ is the mass density and $m_p$ is the proton mass.  Data
gaps in OMNI are filled by linear interpolation for gaps
$\leq$3~hours and flagged for longer gaps; hours with flagged OMNI
data are excluded from training and evaluation.

All inputs used here are archival products: the OMNI series is the final,
cross-calibrated one, assembled days to months after the fact; Kp is a three-hour
index, so at any issue time only the most recently published value is available; and
the cloud fields are ERA5 reanalysis rather than forecasts.  The evaluation that
follows is accordingly retrospective, and Section~\ref{sec:cascade_results}
quantifies the cost of the Kp latency.

\subsection{Magnetic Coordinate Conversion}
\label{sec:data_coords}

\begin{figure}[ht!]
\centering
\includegraphics[width=\columnwidth]{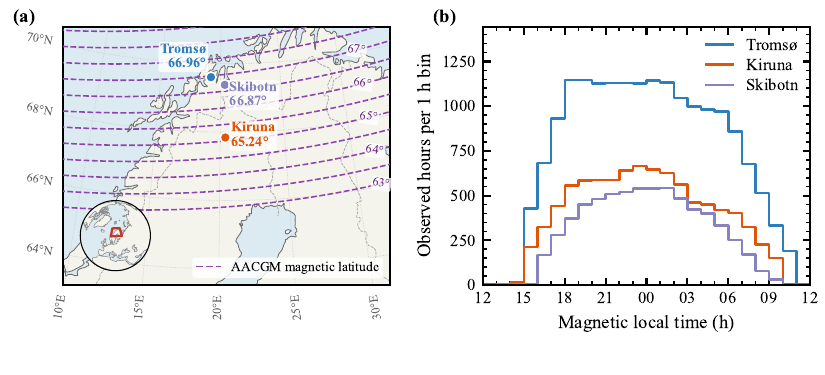}
\caption{(a)~Location of the three primary observation sites.  Dashed contours are
  AACGM-v2 magnetic latitude at 110~km, computed for 2018 with the
  aacgmv2 package \cite{shepherd2014aacgm}; labels give each site's mean
  magnetic latitude over its own observing period; the inset marks the mapped
  region.  (b)~Observed hours per magnetic local time at each site.  Only 208~h
  in total fall within two hours of magnetic noon, because the cameras operate
  only under dark skies.}
\label{fig:aacgm_grid}
\end{figure}

Geographic station coordinates are converted to AACGM-v2
coordinates \cite{shepherd2014aacgm}
at 110~km altitude using the aacgmv2 Python package.  AACGM
coordinates provide physically meaningful MLAT and MLT that
correctly place each station relative
to the auroral oval.  Figure~\ref{fig:aacgm_grid}(a) shows the magnetic
latitude contours over Fennoscandia, which are inclined to the geographic
parallels so that a given magnetic latitude shifts by up to $\sim$10$^{\circ}$ of
longitude across the map.  The contours are drawn for a
single epoch; MLAT and MLT for each station-hour are computed for the corresponding
time.
These coordinates are used both as model inputs and to compute
station distance from the auroral oval (Section~\ref{sec:features}).

\subsection{Atmospheric and Ephemeris Data}
\label{sec:data_atmos}

\begin{figure}[ht!]
\centering
\includegraphics[width=\columnwidth]{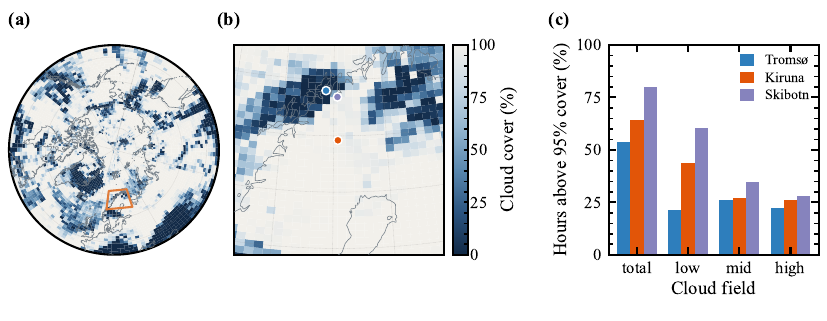}
\caption{Observing-condition data from the Open-Meteo ERA5-based reanalysis
  archive \cite{open_meteo}.  (a)~Total cloud cover over the Northern Hemisphere on
  2019 October 26 at 19:00~UT, on a
  $1.0^{\circ}\times2.5^{\circ}$ grid.  (b)~The same hour over Fennoscandia at
  $0.25^{\circ}\times0.75^{\circ}$, the resolution of the archive; the box in (a)
  marks the region and neither field is interpolated.  Markers give the three sites,
  colored as in (c).  (c)~Share of each site's observed hours in which cover exceeded
  95\%, for the total field and for each layer, over each site's aligned record
  ($n = 17{,}698$, 8{,}885 and 6{,}577 hours).}
\label{fig:cloud_cover}
\end{figure}

For Stage~2 we require local observing-condition data.  Hourly total
and layered cloud cover, the last resolved into low-altitude ($<$2~km),
mid-altitude (2--6~km) and high-altitude ($>$6~km) layers, are obtained from the
Open-Meteo ERA5-based reanalysis archive \cite{open_meteo}.  The
reanalysis is provided on a $0.25^{\circ}$ latitude--longitude grid (about
28~km at these latitudes); values at each station's geographic coordinates are taken
from the nearest grid point.  This provides physically
consistent cloud fields at all locations, enabling hemispheric map
generation in Section~\ref{sec:global_method}.  Figure~\ref{fig:cloud_cover}
shows the field for one hour at both grid resolutions, together with how the total
and layered fields are distributed at the three sites.  Total cover is saturated at
or near 100\% for much of the record, exceeding 95\% in 54, 64 and 80\% of the
observed hours at Troms\o, Kiruna and Skibotn, whereas the individual layers do so
far less often; most of the usable variation therefore lies in the
layers.

Lunar phase and illumination percentage are computed from ephemeris
data using the ephem Python package, providing values for
each station-hour; their effect on the unobscured-classification
rate is quantified in
Section~\ref{sec:stage2_results}.

\section{Label Construction}
\label{sec:labels}

The two-stage framework requires two distinct binary labels for each
station-hour, constructed from the hourly class fractions
introduced in Section~\ref{sec:data_aurora}.  We denote by
$f_{\rm arc}$, $f_{\rm disc}$, $f_{\rm diff}$, $f_{\rm ac}$, and
$f_{\rm ab}$ the hourly mean of the classifier's per-image probability
for the corresponding class, expressed in percent, so that the eight class
fractions of an hour sum to 100.

\subsection{Stage~1 Label: Aurora Identifiable in the Images}
\label{sec:label_s1}

Stage~1 targets aurora occurrence over the station: an hour is labeled
positive if aurora is identifiable in that hour's images, whether against a dark
sky, through cloud (AC), or against a background brightened by moonlight
or sunlight (AB):
\begin{linenomath*}
\begin{equation}
  y_{\rm occ} = \begin{cases}
    1, & \text{if } f_{\rm arc} + f_{\rm disc} + f_{\rm diff}
         + f_{\rm ac} + f_{\rm ab} > 50\%  \\
    0, & \text{otherwise.}
  \end{cases}
  \label{eq:label_occ}
\end{equation}
\end{linenomath*}
The 50\% threshold in Equation~\ref{eq:label_occ} identifies hours in which the auroral classes carry
more than half of the classifier's probability mass, so that
the hour is unambiguously dominated by auroral activity rather than
by clear or cloudy sky without aurora.  The critical distinction from
prior work is the inclusion of
$f_{\rm ac}$ and $f_{\rm ab}$ in the positive class.  Without this,
hours with active aurora obscured by cloud would be labeled
negative, producing an incorrect training signal.

Under optically thick cloud, an aurora that is present is classified as
Cloudy and enters the negative class, so the label rate lies below the
occurrence rate: on dark nightside hours at Troms\o\ the positive rate at fixed Kp
is about 36\% of that on clear-sky hours.  Because the Stage~1 inputs contain no
weather variable, this shortfall enters as label noise nearly independent of the
inputs, depressing the predicted probabilities without reordering them; Stage~1
output is therefore a rank-preserving but downward-biased proxy for occurrence, and
Section~\ref{sec:global_method} recalibrates its level on clear-sky
hours.

\subsection{Stage~2 Label: Unobscured Auroral Image Class}
\label{sec:label_s2}

Among hours in which aurora was identifiable ($y_{\rm occ}=1$), Stage~2
labels whether the hour is dominated by the unobscured auroral classes.  In
\citeA{nanjo2022automated}, Arc, Discrete and Diffuse
are auroral images that are not contaminated by cloud or by a bright
background:
\begin{linenomath*}
\begin{equation}
  y_{\rm obs} = \begin{cases}
    1, & \text{if } f_{\rm arc} + f_{\rm disc} + f_{\rm diff} > 80\%
         \text{ and } y_{\rm occ} = 1  \\
    0, & \text{if } y_{\rm occ} = 1 \text{ and } f_{\rm arc}
         + f_{\rm disc} + f_{\rm diff} \leq 80\%  \\
    \text{undefined}, & \text{if } y_{\rm occ} = 0.
  \end{cases}
  \label{eq:label_obs}
\end{equation}
\end{linenomath*}
The 80\% threshold follows \citeA{nanjo2022automated}, who define auroral
appearance as an image whose summed auroral-class probability exceeds
80\%.  The threshold requires that the unobscured auroral classes
account for most of the hour's probability
mass, excluding hours
dominated by cloud- or brightness-affected aurora
(AC/AB).  This label is defined only
for hours in which aurora was identifiable, so
Stage~2 models the conditional probability of an unobscured classification given
that aurora was identified, rather than re-learning
occurrence patterns.
Throughout this paper we use \emph{visibility} for the image-based quantity
that the two labels define jointly, $y_{\rm occ} = 1$ and $y_{\rm obs} = 1$, an
hour dominated by unobscured auroral imagery (Equation~\ref{eq:label_obs}); it
serves as a proxy for whether the aurora in the images would be visible to an
observer at the site.

\subsection{Data Statistics}
\label{sec:label_stats}

Table~\ref{tab:data_stats} summarizes the dataset composition and
Stage~1 positive fraction $f_+$ across all stations and splits.  The
declining positive rate from
training (14.5\%) to test (8.4\%) at Troms\o{} reflects the Solar
Cycle~24/25 minimum (2019--2020), providing a stringent test of
model robustness under reduced geomagnetic activity.  Kiruna's higher
positive rate (23.9\%) reflects Solar Cycle~25 rising-phase activity
during 2020--2023.  The multi-station training pool holds 16,592 of the
16,771 hours the two training periods contain, the difference being the first six
hours of each continuous segment, which cannot carry the 6-hour rolling
features.

\begin{table*}
\centering
\caption{Dataset composition by site and split.
  $f_+$ gives the Stage~1 (image-identifiable aurora) positive fraction.
  \label{tab:data_stats}}
\small
\begin{tabular}{llccl}
\toprule
Site & Period & $N$ (hr) & $f_+$ (\%) & Role \\
\midrule
Troms\o & 2015--2017 & 8,644  & 14.5 & Train \\
Troms\o & 2018       & 2,860  & 12.5 & Validation \\
Troms\o & 2019--2020 & 6,144  &  8.4 & Test \\
Kiruna  & 2020--2023 & 8,127  & 23.9 & Train (multi-station) \\
Kiruna  & 2024       & 758    & 26.0 & Independent test \\
Skibotn & 2022--2025 & 6,444  & 24.3 & Independent (withheld) \\
\midrule
\multicolumn{2}{l}{Multi-station train total} & 16,592 & 19.0 & \\
\bottomrule
\end{tabular}
\end{table*}

\section{Methods}
\label{sec:methods}

\subsection{Two-Stage Cascade Architecture}
\label{sec:cascade_arch}

Treating occurrence and observability as a single prediction problem
introduces systematic confounders: a model trained on visibility
labels would learn that certain space weather conditions ``produce no
aurora'' when in fact they produce aurora that is simply obscured.

Space weather and tropospheric weather are governed by different
physical systems, so each factor in Equation~\ref{eq:cascade} can be
estimated from its own inputs and improved without the
other.  This architecture yields three
operational advantages: first, Stage~1 uses only globally available
physics inputs, so it can be evaluated at any geographic coordinate, though it is
validated only across the magnetic latitude range the stations
span;
second, each stage can be optimized and updated independently, so that
improved cloud forecasts directly improve Stage~2; third, treating
AC and AB as positive recovers many of the hours a visibility
label would discard, though those under optically thick cloud remain lost
(Section~\ref{sec:label_s1}).

\subsection{Feature Engineering}
\label{sec:features}

\subsubsection{Stage~1: Physics-Only Features}

Stage~1 draws exclusively on globally available physics-driven
features from the data sources described in Section~\ref{sec:data}
(OMNI, AACGM-v2 coordinates): 43~features in five groups.
Table~\ref{tab:features} summarizes them; we detail each below.

\begin{table*}
\centering
\caption{Feature groups for Stage~1.  Features marked $\star$ are newly
  introduced.
  \label{tab:features}}
\small
\begin{tabularx}{\textwidth}{llc>{\raggedright\arraybackslash}X}
\toprule
Group & Abbreviation & $N$ & Physical motivation \\
\midrule
(a) Base space weather + coords & SW+coords & 10 & Instantaneous solar wind drivers and observer position \\
(b) Temporal lags \& rolling stats & Temporal & 14 & Magnetospheric energy storage (1--6~hr response delay) \\
(c) Derived physics & Physics & 8 & Nonlinear coupling ($\varepsilon$, clock angle, $B_z$ duration) \\
(d) Interaction terms$^\star$ & Interact & 3 & Kp$\times$nightside, nightside and storm indicators \\
(e) Multi-station features$^\star$ & Multi & 8 & Oval distance, seasonal/equinox encoding (generalizes across sites) \\
\bottomrule
\end{tabularx}
\end{table*}

(a) Base space weather (10 features):  Nine solar wind and
geomagnetic parameters from OMNI: Kp, $B_z$, $B_y$, $B_x$, $V_{sw}$,
$n_p$, Dst, HPI and $P_{\rm dyn}$ (Equation~\ref{eq:pdyn}), plus
MLT at the observation site.

(b) Temporal lag and rolling features (14 features):
Aurora responds to solar wind conditions, as measured at the L1
Lagrange point by DSCOVR/ACE and time-shifted to the bow shock nose
in the OMNI dataset \cite{king2005omni}, with a 1--6~hour delay
due to magnetospheric energy storage and release
\cite{akasofu1964auroral}.  We construct lag and rolling statistics
for Kp (3h lag, 3h mean, 6h max, 1h difference), $B_z$
(1h lag, 3h mean, 3h/6h minima),
Dst (1h and 3h differences as storm onset indicators), $P_{\rm dyn}$
(1h lag, 3h mean), and MLT cyclic encodings
($\sin(2\pi \cdot {\rm MLT}/24)$, $\cos(2\pi \cdot {\rm MLT}/24)$).

(c) Derived physics features (8 features):
These capture nonlinear magnetospheric dynamics.  They are the $B_z$ southward
duration (consecutive hours of $B_z < 0$, a substorm energy-loading indicator); the
$B_z$ southward magnitude; the transverse IMF $B_T = (B_y^2 + B_z^2)^{1/2}$; the
IMF clock angle $\theta_c = \arctan(B_y/B_z)$, encoded as $\sin\theta_c$ and
$\cos\theta_c$; the Akasofu $\varepsilon$ parameter,
$V_{sw} B_T^2 \sin^4(\theta_c/2)$ \cite{akasofu1981energy}; the cumulative 4-hour sum of the
Newell coupling function \cite{newell2007},
\begin{linenomath*}
\begin{equation}
  \frac{dN}{dt} = V_{sw}^{4/3} B_T^{2/3}
    \sin^{8/3}\!\left(\frac{\theta_c}{2}\right),
  \label{eq:newell}
\end{equation}
\end{linenomath*}
which quantifies the rate of magnetic flux transfer at the dayside magnetopause
(Equation~\ref{eq:newell}); and the Dst recovery indicator
(Dst~$< -20$~nT and $\Delta$Dst$_{1h} > 0$).

(d) Interaction and indicator features (3 features):
The nightside indicator (MLT~$\geq 20$ or $\leq 4$~h), the
Kp$\times$nightside interaction, and a storm indicator (Dst~$< -30$~nT).  The
Kp$\times$nightside feature captures the fact that at the magnetic
latitudes and activity levels our stations sample, identifiable aurora is
overwhelmingly a nightside occurrence.  Dayside aurora is routinely observed at
higher-latitude sites such as Svalbard, and our cameras record only under dark
skies, so this indicator reflects the stations' latitude range and observing
schedule as well as oval physics.

(e) Multi-station generalization features (8 features):
These features enable the model to generalize across sites with
different magnetic latitudes and solar cycle phases.  Seasonal phase
encoding based on day of year (DOY)
($\sin(2\pi \cdot {\rm DOY}/365.25)$,
$\cos(2\pi \cdot {\rm DOY}/365.25)$) and equinox proximity encoding
($\sin(4\pi \cdot {\rm DOY}/365.25)$,
$\cos(4\pi \cdot {\rm DOY}/365.25)$) capture the Russell--McPherron
semiannual variation \cite{russell1973semiannual}.  The normalized
auroral oval distance
\begin{linenomath*}
\begin{equation}
  d_{\rm oval} = \frac{{\rm MLAT} - c_{\rm oval}({\rm Kp, MLT})}
    {\sigma_{\rm oval}({\rm Kp, MLT})},
  \label{eq:oval_dist}
\end{equation}
\end{linenomath*}
where $c_{\rm oval}$ and $\sigma_{\rm oval}$ are derived from the
Feldstein auroral oval parameterization
\cite{feldstein1967auroral,holzworth1975auroral}, encodes each
station's position relative to the auroral oval center, the key
quantity governing aurora occurrence probability at different latitudes.
The 6-hour Kp rolling mean and standard deviation in this group
capture longer-term trend and variability for multi-station
generalization, complementing the shorter lags in group (b).

\subsubsection{Stage~2: Observing-Condition Features}

Stage~2 uses 15 features encoding local atmospheric and observational
conditions, requiring no ground-based aurora data.  Ten of them derive
from the cloud and lunar data described in
Section~\ref{sec:data_atmos} (Open-Meteo ERA5 cloud cover, and lunar
illumination and elevation from ephem).  The remaining five (the two MLT
encodings, the two sector indicators and Kp) describe the aurora rather than
the sky.  They are retained because brightness affects whether an aurora
survives marginal conditions as an unobscured classification.  Stage~2 is
therefore not a pure observing-condition model: the cascade separates the
dominant drivers of the two conditions, not the conditions themselves.  Table~\ref{tab:features_s2}
summarizes the six feature groups; we describe each below.

(a) Raw atmospheric inputs (6 features):
Four cloud cover percentages (total, low-, mid-, high-altitude,
each 0--100\%) directly from the ERA5 reanalysis, plus illumination
percentage (0--100\%) and lunar elevation (degrees above the horizon,
geometric, computed at zero atmospheric pressure) from
ephem.  Lunar phase is not used: illumination percentage is a
deterministic function of phase (the two agree to within 0.06 percentage points),
so including both would carry one quantity twice.

(b) Derived cloud features (1 feature):
Writing $f_{\rm cloud} = C_{\rm total}/100$ and analogously for the low,
mid and high layers, the one derived quantity is a weighted cloud opacity index,
\begin{linenomath*}
\begin{equation}
  O_{\rm cloud} = 1.0\,f_{\rm low} + 0.7\,f_{\rm mid}
    + 0.3\,f_{\rm high},
  \label{eq:cloud_opacity}
\end{equation}
\end{linenomath*}
which accounts for the altitude-dependent optical depth of each cloud layer
(low clouds block aurora light most effectively).

(c) Derived lunar features (2 features):
A sky brightness index
\begin{linenomath*}
\begin{equation}
  B_{\rm sky} = f_{\rm illum} \times \max(\sin h_{\rm moon},\, 0),
  \qquad f_{\rm illum} = \text{illumination\_pct}/100,
  \label{eq:sky_brightness}
\end{equation}
\end{linenomath*}
where $h_{\rm moon}$ is lunar elevation.  Moonlight falling on a horizontal
surface scales as the illuminated fraction times the sine of elevation and vanishes
below the horizon; the expression has no free parameters.  The second feature is a binary high-illumination flag ($= 1$ when
illumination $> 70\%$ \emph{and} the Moon is above the horizon), capturing
the threshold above which faint aurora is no longer separable from the
background (Section~\ref{sec:stage2_results};
Figure~\ref{fig:moon_effect}).

(d) Cloud--moon interaction (1 feature):
The product $f_{\rm cloud} \times f_{\rm illum}$, encoding the joint
effect of cloud cover and lunar brightness on observing conditions.

(e) MLT positional features (4 features):
Cyclic encoding $\sin(2\pi\cdot{\rm MLT}/24)$ and
$\cos(2\pi\cdot{\rm MLT}/24)$, plus pre-midnight
(MLT~$\geq 20$h or MLT~$= 0$h) and post-midnight
(0~$<$~MLT~$\leq 6$h) sector indicators, encoding the diurnal
variation of aurora morphology and brightness.

(f) Aurora intensity proxy (1 feature):
Kp index, which, conditional on aurora being identifiable, correlates with
aurora brightness and improves detection through marginal cloud
conditions.

\begin{table*}
\centering
\caption{Stage~2 observing-condition feature groups.
  \label{tab:features_s2}}
\small
\begin{tabularx}{\textwidth}{llc>{\raggedright\arraybackslash}X}
\toprule
Group & Label & $N$ & Features \\
\midrule
(a) Raw inputs & Raw & 6 & Cloud cover (total, low, mid, high), illumination~\%, lunar elevation \\
(b) Cloud derived & Cloud & 1 & $O_{\rm cloud}$ (Equation~\ref{eq:cloud_opacity}) \\
(c) Lunar derived & Moon & 2 & $B_{\rm sky}$ (Equation~\ref{eq:sky_brightness}), high-illumination flag \\
(d) Cloud--moon & Inter & 1 & $f_{\rm cloud} \times f_{\rm illum}$ \\
(e) MLT position & MLT & 4 & $\sin$/$\cos$(MLT), pre-midnight, post-midnight indicators \\
(f) Aurora intensity & Kp & 1 & Kp (brightness proxy) \\
\bottomrule
\end{tabularx}
\end{table*}

\subsection{Stage~1: XGBoost Occurrence Model}
\label{sec:stage1_model}

We use XGBoost \cite{chen2016xgboost} as the Stage~1 classifier.
Gradient-boosted trees are preferred over deep learning for this
dataset for four reasons: (1)~strong empirical performance on tabular data at
moderate sample sizes ($\sim$16{,}600 training hours); (2)~native
handling of missing values; (3)~interpretable feature importances;
and (4)~computational efficiency enabling extensive hyperparameter
search.  We apply Optuna \cite{akiba2019optuna} with 300 trials using
Tree-structured Parzen Estimator (TPE) sampling; the objective is the joint
validation criterion of Section~\ref{sec:stage2_model}.  Key search ranges:
number of trees [50, 2000], maximum depth [3, 10], learning rate [0.003, 0.3]
(log scale), row subsample [0.5, 1.0], column subsample [0.3, 1.0], minimum child
weight [1, 30], minimum split loss [0, 5], and positive-class weight
[1.0, $n_-/n_+$].  Early stopping with
patience~$= 150$ rounds (on validation log loss) prevents overfitting.
The search selected 723 trees of depth 10 at a learning rate of 0.011, with
row subsample 0.843, column subsample 0.940, minimum child weight 19, minimum
split loss 0.49 and positive-class weight 5.99.

All splits are strictly time-based to prevent temporal leakage.
Training and validation use the same station--period splits and
warm-up exclusion as in Table~\ref{tab:data_stats}
(Section~\ref{sec:label_stats}).  The multi-station training pool
combines Troms\o\ 2015--2017 with Kiruna 2020--2023; validation uses
Troms\o\ 2018; the test set uses Troms\o\ 2019--2020; Skibotn
2022--2025 is withheld entirely as an independent site.

Raw XGBoost probabilities are calibrated by Platt scaling
\cite{platt1999probabilistic}, fitted on the validation set and applied to test
and independent sets without refitting; the sigmoid is monotone and smooth, so the
calibrated probabilities take distinct values on which the reliability diagram of
Section~\ref{sec:cascade_results} can be drawn.  We compute two
operating points from the validation-set precision--recall curve: (1)~the
F1-optimal threshold, maximizing the balance between precision and recall;
(2)~the F0.5-optimal threshold ($\beta = 0.5$), favoring precision
over recall, so that downstream applications can select their
preferred operating point.

\subsection{Stage~2: Observing Conditions Model}
\label{sec:stage2_model}

Stage~2 is trained exclusively on aurora-occurring hours
($y_{\rm occ} = 1$), modeling the conditional probability
$P(\text{unobscured} \mid \text{identifiable})$.  We compared
three approaches: (a)~a physical multiplicative
model $(1 - f_{\rm cloud})(1 - f_{\rm illum})$, where $f_{\rm cloud}$
and $f_{\rm illum}$ denote cloud cover and lunar illumination
fractions (0--1), respectively; (b)~logistic regression (LR) with
L2 regularization; and (c)~XGBoost.
For this comparison hyperparameters were tuned by five-fold
cross-validation within the training set, and the choice among the three
families was made on the validation set; the test set and the independent sites
took no part in the selection and were evaluated once, after it.  Where two
candidates fell within each other's bootstrap confidence interval on validation,
we preferred the more interpretable model.  In this comparison, before the joint tuning described below, validation ROC-AUC
was 0.638 for the physical model, 0.831 for LR and 0.849 for XGBoost, and this
validation set does not resolve the difference between the latter two (95\% confidence interval on
the difference $-0.003$ to $+0.039$, $n = 358$).  XGBoost reaches 0.872 in
cross-validation within the training set but 0.849 on validation, against 0.835
and 0.831 for LR, indicating that the tree ensemble overfits the small
conditional sample.  We therefore select LR followed by isotonic calibration
\cite{zadrozny2002calibration}; its standardized coefficients can be read directly
as the effect of each observing-condition variable
(\ref{app:shap}).  Calibration is fitted on the validation set
and applied to test and independent sets without refitting.

With the families fixed, the final hyperparameters of both stages, Stage~1's eight XGBoost parameters and Stage~2's regularization strength ($C =
4.09$), are set together by a 300-trial Optuna search whose objective is the sum of the
cascade's log loss and Stage~1's occurrence log loss on the validation set.  A
joint search removes the asymmetry of tuning each stage against its own
intermediate label, and log loss rather than ROC-AUC is the criterion because
the cascade's validation AUC is nearly flat in $C$
(0.960--0.961 across four orders of magnitude), so a ranking objective selects
on noise, whereas log loss is a strictly proper score with a clear minimum that
constrains calibration and discrimination together.

\subsection{Hemispheric Occurrence Field}
\label{sec:global_method}

Three stations within two degrees of magnetic latitude cannot, on their own,
determine how occurrence varies across the hemisphere.  The construction adopted
here divides the problem.  Each station enters Stage~1 by its position relative to
the auroral oval rather than by its identity (Equation~\ref{eq:oval_dist}), so
Stage~1 supplies the activity level at the oval for a given hour; an established
parameterization of the oval supplies where that activity lies; and paired
observations at the stations test the direction of the latitude dependence.  The
field is
\begin{linenomath*}
\begin{equation}
  P_{\rm occ}(\Lambda, \tau, t) = A(t)\,
    \exp\!\left[-\frac{1}{2}\left(\frac{\Lambda - c(\tau, {\rm Kp})}
                                       {w(\tau, {\rm Kp})}\right)^{\!2}\right],
  \label{eq:global_occ}
\end{equation}
\end{linenomath*}
for magnetic latitude $\Lambda$ and magnetic local time $\tau$, where $c$ and $w$
are the center and half-width of the Feldstein auroral oval
\cite{feldstein1967auroral,holzworth1975auroral} and $A(t)$ is the occurrence
probability at the oval center for the hour $t$.

The division is deliberate because the latitude dependence in
Equation~\ref{eq:global_occ} cannot be learned from a single site: at a fixed site the normalized oval distance of
Equation~\ref{eq:oval_dist} is a deterministic function of Kp and magnetic local
time, so a latitude profile fitted to single-site data cannot be separated from the
activity response.  What the data do support is a paired comparison.  Hours observed
simultaneously at two sites share the same solar wind and differ only in oval
distance, and a conditional-logistic estimate on the 878 such pairs in which exactly
one site records aurora, with a free intercept per station pair so that
inter-camera sensitivity differences cannot masquerade as a latitude effect,
gives $\kappa = -1.24$ on the squared oval distance, 95\% interval
$[-2.15, -0.31]$, negative in every bootstrap replicate.  The sign, occurrence
falling away from the oval center, is therefore established from our own
network without reference to the parameterization.

To test that sign with substantial poleward leverage, we added a fourth site: 4{,}500
hourly frames from the ArcLight all-sky camera at Ny-\r{A}lesund
\cite{johnsen2021arclight,arclight_nird} (MLAT $76.8^{\circ}$, winters
2022--2025) were labeled by four human annotators under a three-way protocol in
which fully overcast frames are excluded rather than counted as negatives, so
that the cloud censoring of Section~\ref{sec:label_s1} is not re-created in the
new data; the labels are released with the data package (Open Research).  Adding the
Ny-\r{A}lesund pairs raises the median oval-distance contrast of a discordant
pair from 0.11 to 2.9, and the estimate becomes $\kappa = -1.94$, 95\% interval
$[-2.98, -0.25]$, negative in 99\% of replicates and stable under removal of any
single annotator.  The implied half-width, 0.51 with interval $[0.41, 1.27]$,
still spans the parameterization's 1.0, so
Equation~\ref{eq:global_occ} retains the parameterization's width.

The same labels give the field its first test outside the training latitudes:
on the 1{,}466 usable nightside hours at Ny-\r{A}lesund, Equation~\ref{eq:global_occ}
predicts a mean occurrence of 0.120 against an observed 0.141, so its level
survives an extrapolation of nine degrees poleward.

An independent measurement system corroborates the direction of the falloff.
Applying the same paired estimator to calibrated 557.7~nm intensities from three
photometric stations of a different network \cite{vandekamp2026intensities}
(Kilpisj\"arvi, Abisko and Sodankyl\"a, MLAT $66.4^{\circ}$, $65.7^{\circ}$ and
$64.5^{\circ}$; 2012--2024), on hours with lunar illumination below 40\% and at a
bright-aurora threshold of 2~kR, reproduces the negative sign in every bootstrap
replicate, with the expected ordering that brighter aurora is confined more
tightly to the oval.
Stage~1 supplies $A(t)$, the probability that aurora occurs in the hour.

The amplitude $A(t)$ is Stage~1 recalibrated on clear-sky hours, those the reanalysis
places under less than one-third total cloud cover
(Section~\ref{sec:global_results}, Figure~\ref{fig:kp_response}), because
$P_{\rm occ}$ is multiplied by Stage~2 in the next step and the occurrence label is
not independent of cloud.  The eight camera classes sum to 100\%, so the occurrence label
$y_{\rm occ}$, defined by the five aurora classes exceeding 50\%, is
equivalent to the three non-aurora classes falling below 50\%: an hour more than half
cloudy cannot carry a positive label.  As a camera-based check, among nightside hours above Kp~$= 2$ in which
the camera is not majority cloud, the label is positive for 97.0\% ($n = 1{,}484$),
so at moderate and higher activity the all-hours occurrence rate measures sky
availability as much as it measures aurora.

The consequence is a level rather than a shape.  Fitting the effective
half-width of the occurrence profile separately under each cloud condition gives
0.47 to 0.51 with overlapping intervals, and mean cloud cover is flat across oval
distance (79.8--81.4\% across quintiles), so censoring does not generate the
latitude profile.  It does move the level: above Kp~$= 2$ the clear-sky occurrence rate agrees
between Troms\o\ and Kiruna while the all-hours rate does not
(Section~\ref{sec:global_results}, Figure~\ref{fig:kp_response}), so the clear-sky
rate is the site-invariant quantity and the amplitude is fitted on the clear subset.
Recalibrated there, $A(t)$ is an occurrence probability that does not already carry
cloud, so multiplying by Stage~2 counts each factor once.

The field thus multiplies a learned temporal factor by an assumed spatial
one; Section~\ref{sec:limitations} states the range of latitude and activity over
which it has been checked.

\section{Results}
\label{sec:results}

All results in this section are retrospective: the models are scored on archival
inputs at the target hour, so the scores can be regarded as an upper bound on what a
real-time implementation would achieve (\ref{app:operational}).

\subsection{Observing-Condition Examples}
\label{sec:condition_examples}

\begin{figure*}[ht!]
\centering
\includegraphics[width=\columnwidth]{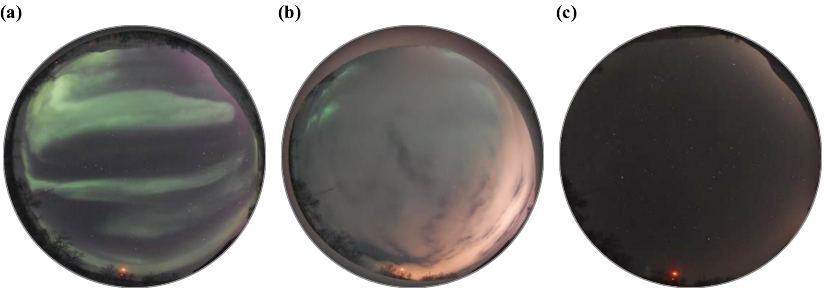}
\caption{Three hours from the Troms\o\ 2019--2020 test period,
  spanning the two ways a night fails to deliver a visible aurora.
  (a)~2019 February 4, 23:00~UT: aurora occurred and was recorded.
  (b)~2019 February 7, 20:00~UT: aurora occurred under 97\% cloud.
  (c)~2019 February 5, 04:00~UT: clear sky with no aurora.
  Images are from the Troms\o\ all-sky imager \cite{nanjo2022automated},
  masked to the field of view and shown with a common brightness
  stretch; the cases are discussed in
  Section~\ref{sec:condition_examples}.}
\label{fig:case_examples}
\end{figure*}

Three hours from the Troms\o\ 2019--2020 test period span the two
ways a night fails to deliver a visible aurora
(Figure~\ref{fig:case_examples}); we write $P_{\rm occ}$ and $P_{\rm clear}$ for
the two factors of Equation~\ref{eq:cascade} and $P_{\rm vis}$ for their product.
All three fall within three days of one another at the same site, with the Moon
less than 7\% illuminated, so season, instrument and moonlight are held fixed.

\textbf{Case~A} (aurora occurs, sky clear).  2019 February 4,
23:00~UT.  Kp~$= 2.3$, $B_z = -0.7$~nT, Dst~$= -21$~nT, cloud cover
$= 0$\%, lunar illumination~$= 0.3$\%.  Predictions:
$P_{\rm occ} = 0.59$, $P_{\rm clear} = 0.92$, $P_{\rm vis} = 0.54$.
Observed: aurora visible
(Discrete 57\%, Arc 36\%).  Both terms are favorable
and the aurora was recorded.

\textbf{Case~B} (aurora occurs, sky obscured).  2019 February 7,
20:00~UT.  Kp~$= 1.0$, $B_z = -3.4$~nT, Dst~$= -12$~nT, cloud cover
$= 97$\%, with the low and mid layers both near 90\%, lunar
illumination~$= 7.0$\%.  Predictions: $P_{\rm occ} = 0.51$,
$P_{\rm clear} = 0.19$, $P_{\rm vis} = 0.10$.
Observed: aurora through cloud (AC
84\%, Cloudy 16\%).
Aurora is identifiable in the images, and Stage~1
assigns an occurrence probability close to that of Case~A because $B_z$ is more
strongly southward despite the lower Kp; in our labeling
(Section~\ref{sec:label_s1}) this hour is a positive occurrence example for
Stage~1.  Visibility falls by a factor of 5.7 relative to Case~A, almost entirely
through the observing term.

\textbf{Case~C} (no aurora, sky clear).  2019 February 5,
04:00~UT, five hours after Case~A on the same clear night.
Kp~$= 1.3$, $B_z = 0.0$~nT, Dst~$= -14$~nT, cloud cover~$= 0$\%, lunar
illumination~$= 0.3$\%.  Predictions: $P_{\rm occ} = 0.08$,
$P_{\rm clear} = 0.62$, $P_{\rm vis} = 0.05$.
Observed: no aurora (Clear 100\%).
The observing term remains favorable; the reduction comes from the
occurrence term.

Cases~B and C reach a similarly low visibility probability by
opposite routes.  A model trained directly on visibility returns one
number for both and gives no indication of which term is responsible,
whereas the cascade separates them: Case~B is a night to wait out the
cloud, Case~C a night to wait for activity.

\subsection{Stage~1: Aurora Occurrence Prediction}
\label{sec:stage1_results}

The hyperparameters selected by Optuna (Section~\ref{sec:stage1_model})
favor deep trees grown slowly, a depth of 10 at a learning rate of 0.011 over
723 rounds, with strong leaf regularization against over-fitting the
multi-station distribution, and a positive-class weight of 5.99 compensates for
class imbalance, with
Platt-scaling calibration applied after training.

Table~\ref{tab:stage1_results} presents Stage~1 performance across
all evaluation sets.  Because the model excludes all ground-based
observation features, its validation AUC of 0.856 (Troms\o\ 2018) reflects
predictive power from space weather inputs and magnetic coordinates
alone.  The test set positive rate of only 8.4\%
(Solar Cycle~24/25 minimum, 2019--2020) depresses the attainable precision;
nonetheless, ROC-AUC~$= 0.849$ shows that ranking quality is maintained despite
the distribution shift.  Kiruna 2024 (AUC~$= 0.753$) is a new time period at a station whose
earlier years are in the training set, not a new station; Skibotn (AUC~$= 0.678$)
is the only station withheld entirely, and the drop from 0.849 at Troms\o\ is the
cost of moving to a new site.

\begin{table*}
\centering
\caption{Stage~1 aurora occurrence classification performance.
  Splits are Troms\o\ unless stated.  ROC-AUC is threshold-independent;
  precision, recall and F1 use the validation-set F1-optimal threshold
  $t = 0.173$.  The alternative F0.5-optimal operating point,
  $t_{0.5} = 0.378$, is marked in Figure~\ref{fig:pr_curve}.
  \label{tab:stage1_results}}
\small
\begin{tabular}{llcccc}
\toprule
Split & $N$ (pos\%) & Prec & Rec & F1 & ROC-AUC \\
\midrule
Train (Troms\o+Kiruna) & 16{,}592 (19.0\%) & 0.534 & 0.993 & 0.694 & 0.978 [0.975, 0.980] \\
Val 2018               & 2{,}860 (12.5\%)  & 0.347 & 0.726 & 0.469 & 0.856 [0.831, 0.878] \\
Test 2019--20          & 6{,}144 (8.4\%)   & 0.235 & 0.761 & 0.359 & 0.849 [0.829, 0.868] \\
\midrule
Kiruna 2024            & 758 (26.0\%)      & 0.466 & 0.599 & 0.524 & 0.753 [0.693, 0.805] \\
Skibotn 2022--25       & 6{,}444 (24.3\%)  & 0.348 & 0.667 & 0.457 & 0.678 [0.653, 0.704] \\
\bottomrule
\end{tabular}
\end{table*}

\begin{figure}[ht!]
\centering
\includegraphics[width=\columnwidth]{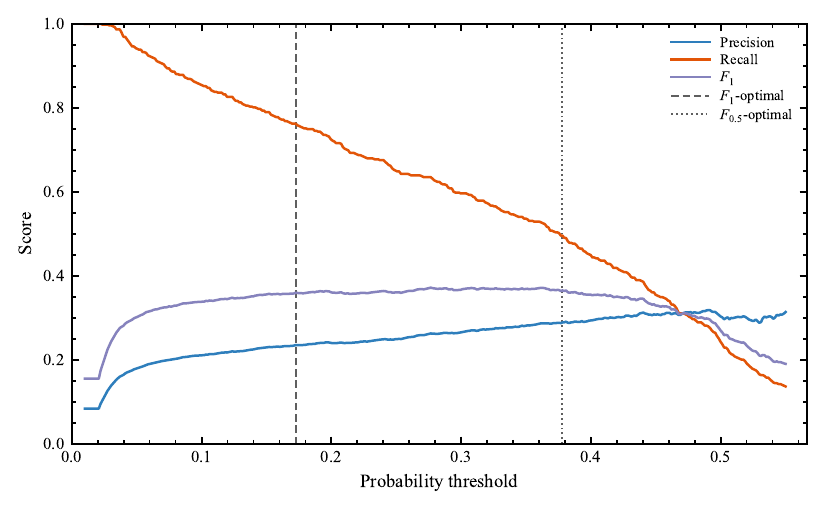}
\caption{Precision (blue), recall (orange) and $F_1$ score (purple) of the
  multi-station Stage~1 model on the Troms\o\ 2019--2020 test set, as functions of
  the probability threshold.  The two vertical lines, named in the legend, mark the
  operating points selected on the validation set; their values are given in the text.  As the threshold
  increases, precision improves modestly while recall drops sharply, reflecting the
  class imbalance (8.4\% positive rate; 518 of 6{,}144 hours).}
\label{fig:pr_curve}
\end{figure}

Figure~\ref{fig:pr_curve} shows the test-set precision, recall, and F1
score as functions of the probability threshold, allowing selection of
application-specific operating points.  The two
validation-set-optimal thresholds, $t = 0.173$ for F1 and
$t_{0.5} = 0.378$ for F0.5, are marked.  For aurora
tourism, the F1 threshold provides high recall (76\%) to minimize
missed aurora nights; the F0.5 threshold trades 26 points of recall for
5 points of precision, and suits users for whom a false alarm costs
more than a missed night.

Feature attributions and the feature-group ablation for Stage~1 are given in
\ref{app:shap}: magnetic local time, the Kp$\times$nightside interaction
and the three-hour mean Kp dominate, carrying 44\% of SHAP attribution, and no
single feature group is indispensable.

\begin{table*}
\centering
\caption{Three-station aurora occurrence validation on nightside
  hours, with the Brier score of Equation~\ref{eq:brier}.  The model is trained on
  Troms\o\ 2015--2017 and Kiruna 2020--2023, so those years are in sample; Skibotn
  is withheld entirely.
  \label{tab:three_station}}
\small
\begin{tabular}{lcccc}
\toprule
Station & MLAT & $N$ (nightside) & AUC & Brier \\
\midrule
Troms\o               & 66.96$^{\circ}$ & 8{,}810 & 0.863 [0.846, 0.878] & 0.113 \\
Kiruna                & 65.24$^{\circ}$ & 4{,}692 & 0.948 [0.936, 0.958] & 0.117 \\
Skibotn                   & 66.87$^{\circ}$ & 3{,}958 & 0.595 [0.559, 0.630] & 0.220 \\
\bottomrule
\end{tabular}
\end{table*}

Table~\ref{tab:three_station} quantifies Stage~1 occurrence performance on the
nightside hours of all three stations.  The Troms\o\ row spans all six
years, four of which the model saw in training or validation; on the held-out
2019--2020 nightside hours alone ($n = 3{,}181$) the AUC is 0.752 [0.718, 0.785],
the number to compare with the other sites.  The Kiruna AUC
($0.948$) is likewise in sample, since Kiruna
2020--2023 was part of the training set; the fully independent Kiruna
2024 set ($n = 758$) yields occurrence AUC~$= 0.753$ and cascade
AUC~$= 0.933$ (Table~\ref{tab:cascade_results}); its occurrence Brier
score of 0.176 is a skill score of $+0.09$ against the base rate of that
set.  Skibotn is the only station withheld from training, so its AUC
($0.595$) is the paper's one measurement of transfer to a new site.  It is well
above chance, consistent with the oval-relative coordinate
(Equation~\ref{eq:oval_dist}, Section~\ref{sec:features}) and the seasonal
encodings carrying across sites, but it is 0.157 below the Troms\o\ value on that site's own
held-out years, so transfer is partial.

\subsection{Stage~2: Observing Conditions Model}
\label{sec:stage2_results}

Table~\ref{tab:stage2_results} presents Stage~2 performance.  Stage~2
is trained on $n = 3{,}200$ aurora-occurring hours (Troms\o\ 2015--2017 and
Kiruna 2020--2023, 31.6\% unobscured) and evaluated on independent
time periods and sites.  Evaluation at the other sites uses all available Kiruna aurora-occurring
hours (2020--2024, in sample for 2020--2023) and, at the station withheld
entirely, all 1{,}587 Skibotn aurora-occurring hours (2022--2025).

\begin{table*}
\centering
\caption{Stage~2 observing-conditions model performance
  (LR + isotonic calibration, trained on aurora-occurring hours only).
  Splits are Troms\o\ unless stated; precision, recall and F1 use the
  validation-set F1-optimal threshold $t = 0.462$.
  Brier score: Equation~\ref{eq:brier}.
  \label{tab:stage2_results}}
\small
\begin{tabular}{llccccc}
\toprule
Split & $N$ (pos\%) & Prec & Rec & F1 & ROC-AUC & Brier \\
\midrule
Train (Troms\o+Kiruna) & 3{,}200 (31.6\%) & 0.565 & 0.819 & 0.669 & 0.836 [0.815, 0.855] & 0.153 \\
Val 2018       & 358 (41.3\%)     & 0.641 & 0.919 & 0.756 & 0.840 [0.787, 0.888] & 0.155 \\
Test 2019--20  & 519 (33.7\%)     & 0.548 & 0.846 & 0.665 & 0.818 [0.767, 0.865] & 0.164 \\
\midrule
Kiruna 2020--24 & 2{,}138 (23.4\%) & 0.498 & 0.840 & 0.625 & 0.853 [0.828, 0.877] & 0.133 \\
Skibotn 2022--25 & 1{,}587 (30.4\%) & 0.540 & 0.629 & 0.581 & 0.750 [0.711, 0.787] & 0.182 \\
\bottomrule
\end{tabular}
\end{table*}

The standardized LR coefficients (\ref{app:shap},
Table~\ref{tab:stage2_coefs}) identify moonlight illuminating the sky, $B_{\rm sky}$,
and mid-level cloud as the strongest obscuring terms: at illumination above
80\% the unobscured-classification rate falls to 8.9\% (Figure~\ref{fig:moon_effect}).
Kiruna, in sample for 2020--2023, reaches AUC 0.853, and Skibotn, withheld
entirely, 0.750 against 0.818 on the Troms\o\ test years, so the
observing-condition relation carries to a new site with a modest loss.

\subsection{Cascade Integration}
\label{sec:cascade_results}

\begin{figure}[ht!]
\centering
\includegraphics[width=\columnwidth]{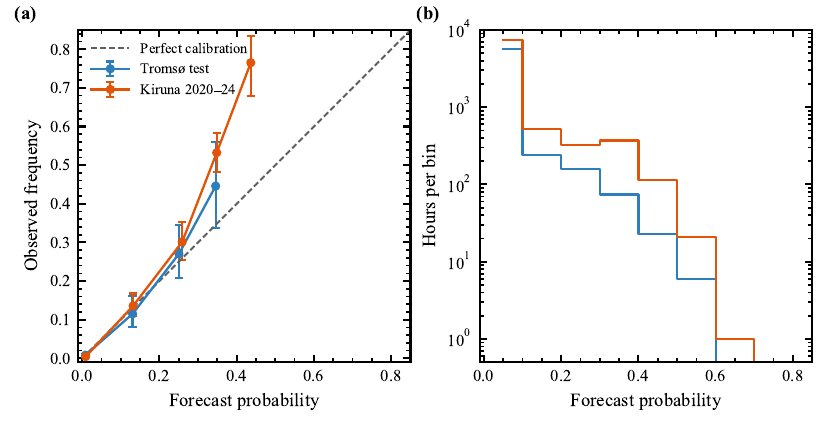}
\caption{Reliability of the cascade probabilities.  (a)~Observed frequency
  against forecast probability, in bins of width 0.1 holding at least 30 hours;
  vertical bars are Wilson 95\% intervals and the dashed line is perfect
  calibration.  At Troms\o\ the forecast lies inside the interval in every bin; at
  Kiruna it falls below the interval in the two highest bins, 0.35 against an
  observed 0.53 and 0.44 against 0.77.  (b)~Hours in the same bins, so each step is
  the count behind the point above it.
  \label{fig:reliability}}
\end{figure}

Table~\ref{tab:cascade_results} presents end-to-end performance for
the cascade target across all hours: the probability that an hour is
dominated by unobscured auroral imagery.  Cascade evaluation is reported for Troms\o\ and Kiruna, the two sites
whose earlier years enter training; Skibotn enters the evaluation of each stage
separately (Sections~\ref{sec:stage1_results} and~\ref{sec:stage2_results}).  Using
Stage~1 \emph{alone} as a visibility proxy (i.e.\ $P_{\rm occ}$
without cloud/moon correction) already achieves AUC~$= 0.863$
(Troms\o) and $0.836$ (Kiruna 2024).  Adding Stage~2 further raises
AUC to 0.958 on Troms\o\ ($+0.095$) and 0.933 on Kiruna 2024 ($+0.097$),
confirming that observing conditions contribute substantial predictive
information beyond occurrence alone.  Both gains are resolved: DeLong's test for correlated ROC curves
\cite{delong1988comparing}, in the fast implementation of \citeA{sun2014fast} and as
applied to geomagnetic forecasts by \citeA{chakraborty2020probabilistic}, gives
$p < 10^{-49}$ at Troms\o\ and
$p = 1.4 \times 10^{-3}$ at Kiruna, after Bonferroni correction for the two site
comparisons.  We assess calibration using the
Brier score \cite{brier1950verification},
\begin{linenomath*}
\begin{equation}
  \mathrm{BS} = \frac{1}{N}\sum_{i=1}^{N}(p_i - o_i)^2,
  \label{eq:brier}
\end{equation}
\end{linenomath*}
where $p_i$ is the predicted probability and $o_i \in \{0,1\}$ is
the observed outcome for sample~$i$.  The Brier score ranges from 0
(perfect) to 1 (worst) and jointly captures two desirable properties:
\emph{discrimination}, whether the model assigns higher probabilities
to positive events, and \emph{calibration}, whether a predicted
probability of, e.g., 30\% corresponds to an observed frequency of
$\approx$30\%.  A small Brier score is not on its own evidence of
calibration when the positive class is rare: on the Troms\o\ test set, where
2.8\% of hours are aurora-visible, a constant forecast at that base rate already
scores 0.028.  Against that reference the cascade's 0.021 is a Brier skill score
of $+0.25$, and $+0.26$ at Kiruna (0.030 against a base-rate reference of 0.040).  Figure~\ref{fig:reliability} shows where the
probabilities stand: in its highest bin the cascade forecasts 0.35 against an
observed 0.45 at Troms\o, where the forecast lies inside the interval, and 0.44
against 0.77 at Kiruna, where it does not, so where it is most confident it
under-forecasts rather than over-forecasts, the safer direction for a planning
product, though at Kiruna by a wide margin.

A single model trained directly on the final target, with the same inputs,
tuning budget and objective, reaches an indistinguishable test AUC (0.964 against
0.958; DeLong $p = 0.083$), and the same holds on the independent Kiruna 2024
hours (0.953 against 0.933, $n = 758$; $p = 0.34$).  The decomposition is
retained for what it affords beyond accuracy.  Each stage can be read on its own:
the observing-conditions coefficients are physical effects
(\ref{app:shap}), and a low forecast can be traced to the factor that
caused it (Section~\ref{sec:condition_examples}).  Each stage can be updated on
its own: the observing-conditions stage carries across sites
(Section~\ref{sec:stage2_results}) and can be driven by a weather forecast
without retraining the space-weather stage.  And the occurrence factor alone
extends to the hemispheric field of Section~\ref{sec:global_method}, which a
single visibility model could not provide.

One of the ways in which the archival inputs differ from what a live system would
receive can be measured directly.  Substituting the most recently published three-hour Kp (the only Kp available
at an issue time) for the concurrent value moves the cascade from
0.958 to 0.952 at Troms\o\ and leaves it at 0.933 at Kiruna~2024; because Kp is a
three-hour index, the most recently published value equals the concurrent one in 69.5\% of
hours.  Of the two remaining differences, the
solar-wind and geomagnetic-index half is measured by replaying the operational
products themselves: over 571 hours at Skibotn the two input chains agree at $r = 0.91$ and
rank the same hours, a paired ROC-AUC difference of $+0.004$ with a 95\% bootstrap interval
of $[-0.011, +0.020]$ (\ref{app:operational}).  The other half, the use of a
reanalysis where an operational system would hold a cloud forecast, cannot be quantified
without an archive of those forecasts and can only reduce skill.  The figures reported here can
therefore still be regarded as an upper bound on what a real-time implementation
would achieve.

\begin{table*}
\centering
\caption{End-to-end cascade evaluation for aurora visibility
  prediction.  The cascade operates on all hours (not just
  aurora-occurring hours).  ``Stage~1 alone'' uses $P(\text{occurring})$
  as a visibility proxy without the cloud/moon correction.
  Best F1 is reported only for the cascade; Stage~1 alone is not
  threshold-tuned for visibility.  Brackets give bootstrap 95\% confidence
  intervals on ROC-AUC (24-hour blocks, 2000 resamples).
  \label{tab:cascade_results}}
\small
\begin{tabular}{llcccc}
\toprule
Site & Model & ROC-AUC & Avg Prec & Brier & Best F1 \\
\midrule
Troms\o\ Test   & Stage~1 alone  & 0.863 [0.834, 0.888] & 0.112 & 0.055 &---\\
Troms\o\ Test   & Cascade (S1$\times$S2) & 0.958 [0.946, 0.968] & 0.449 &
    0.021 & 0.459 \\
\midrule
Kiruna 2024 & Stage~1 alone  & 0.836 [0.758, 0.886] & 0.123 & 0.072 &---\\
Kiruna 2024 & Cascade (S1$\times$S2) & 0.933 [0.846, 0.978] & 0.514 &
    0.030 & 0.539 \\
\bottomrule
\end{tabular}
\end{table*}

Per-year cascade AUC (\ref{app:yearly},
Figure~\ref{fig:yearly_auc}) is consistent across the record.  The three
years held out from both stages (Troms\o\ 2019 and 2020 and Kiruna 2024) give
0.962, 0.946 and 0.933, and each interval covers their pooled value of 0.954
over 6{,}902 hours and 207 events.
Across the whole record no year differs significantly from any
other.
Kiruna spans the early rising phase of Solar Cycle~25 (2020--2021, AUC
0.957 and 0.958) and its approach to maximum (2022--2024, AUC 0.940, 0.960 and
0.933), so the record covers both quiet and active phases.

\subsection{Hemispheric Probability Maps}
\label{sec:global_results}

\begin{figure*}[ht!]
\centering
\includegraphics[width=\textwidth]{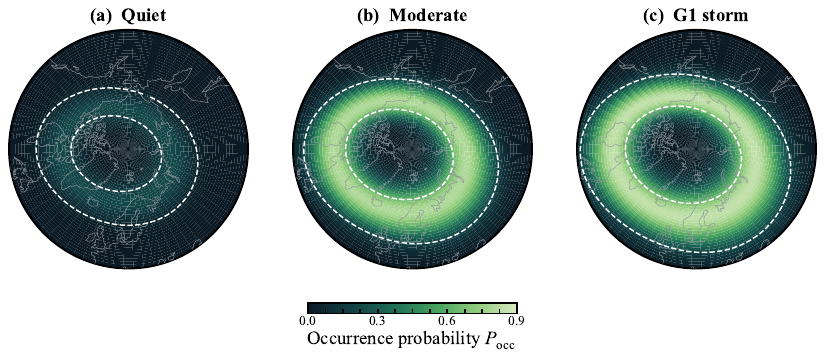}
\caption{Occurrence probability $P_{\rm occ}$ over the Northern Hemisphere
  (north polar stereographic, $45^{\circ}$--$90^{\circ}$N) for three held-out hours
  near magnetic midnight, each with aurora recorded on the ground.
  \textit{(a)}~2020 December~15, 22:00~UT, Kp~$= 1.0$.  \textit{(b)}~2019
  October~26, 22:00~UT, Kp~$= 3.7$.  \textit{(c)}~2019 September~27, 22:00~UT,
  Kp~$= 5.0$.  The amplitude is Stage~1 on the clear-sky basis and the latitude
  structure is the Feldstein parameterization's (Section~\ref{sec:global_method}).
  The three panels share one color scale.  Dashed curves bound the
  oval-distance interval the stations sample ($-1.66$ to $+0.69$); outside them the
  field is extrapolation.}
  \label{fig:global_3panel}
\end{figure*}

Figure~\ref{fig:global_3panel} shows the occurrence field of
Equation~\ref{eq:global_occ} for three held-out hours spanning quiet to storm
conditions, each with its measured solar-wind state rather than a state
synthesized from Kp, and each under a moon less than 6\% illuminated.  The
amplitude, Stage~1 on the clear-sky basis, is $A = 0.22$, $0.80$ and $0.83$; the
ring's position and width are the Feldstein parameterization's.

Across the three hours the mean probability over $45^{\circ}$--$85^{\circ}$N
rises from 0.058 to 0.295 to 0.353, and the fraction of that band above $P = 0.1$
from 26.9\% to 60.1\% to 69.7\%.  Most of that growth is the ring moving rather than
brightening: the amplitude rises from 0.22 to 0.80 between the quiet and moderate hours but
only to 0.83 at storm level, while the nightside peak continues equatorward from
$67^{\circ}$ to $63^{\circ}$ to $61^{\circ}$ MLAT.  That migration is that of the Feldstein
oval midpoint, which the field inherits from the parameterization.

\begin{figure*}[ht!]
\centering
\includegraphics[width=\textwidth]{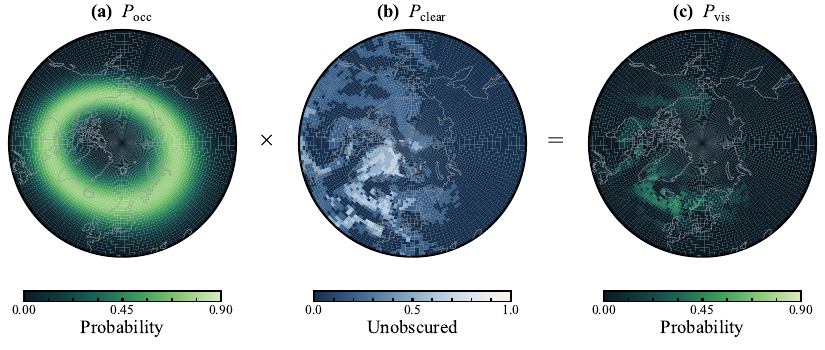}
\caption{The cascade over the Northern Hemisphere for 2024 January~1,
  22:00~UT, an hour in the held-out Kiruna 2024 period (Kp~$= 4.0$, Moon 69.4\%
  illuminated at $11^{\circ}$ elevation).  \textit{(a)}~Occurrence probability
  $P_{\rm occ}$ from Equation~\ref{eq:global_occ}, with amplitude $A = 0.70$
  from Stage~1 on the clear-sky basis.  \textit{(b)}~The Stage~2 model evaluated
  at every grid cell on that cell's own reanalysis cloud cover and lunar elevation.
  \textit{(c)}~Their product, on the color scale of (a).  At the Kiruna cell the
  decomposition reads $0.70 \times 0.20 = 0.14$, and the camera recorded aurora as
  visible that hour.}
\label{fig:cascade_layers}
\end{figure*}

Figure~\ref{fig:cascade_layers} shows the full cascade decomposed over the
hemisphere for a single held-out hour: the occurrence field, the
observing-conditions field evaluated on each cell's own cloud cover and lunar
elevation, and their product, the visibility field.  Stage~2 uses no latitude
variable, and cloud cover and illumination span their full training range, so
evaluating it away from the stations is interpolation; its dayside values are
extrapolation, since Stage~2 is trained almost entirely on nightside hours, and are
shown rather than masked because a camera cannot operate there in any case.

\begin{figure}[ht!]
\centering
\includegraphics[width=\columnwidth]{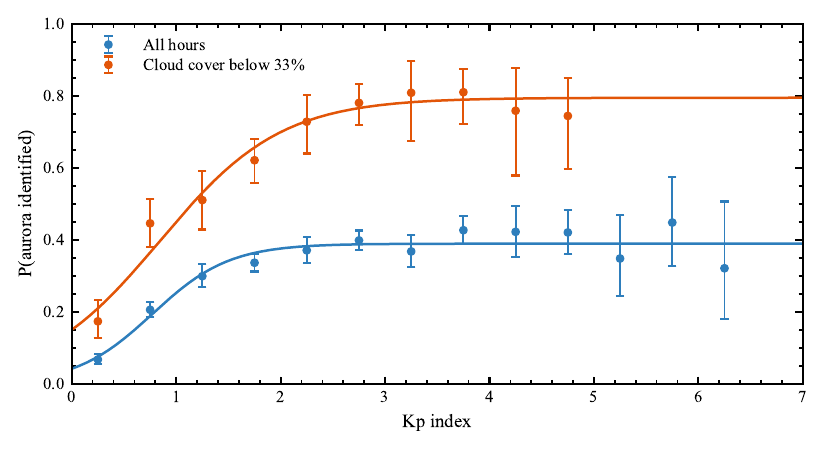}
\caption{Nightside occurrence rate against Kp over the multi-station training
  set (Troms\o\ 2015--2017 and Kiruna 2020--2023, MLT~$\geq 20$ or $\leq 4$), in Kp
  bins of width 0.5 holding at least 20 hours; vertical bars are Wilson 95\%
  intervals and the curves are logistic fits.  The lower series is all hours, the
  upper the 15.7\% of hours the reanalysis places under less than a third cloud
  cover.}
\label{fig:kp_response}
\end{figure}

Figure~\ref{fig:kp_response} is the measurement behind the clear-sky basis.
Both series saturate above Kp~$\approx 2$, so the saturation belongs to the
occurrence rate rather than to the observing conditions, and they differ by a factor
of about two (asymptotes 0.390 against 0.795) because the occurrence label requires
aurora to be identifiable in the images: an hour more than half obscured cannot be
positive.  Above Kp~$= 2$ the clear-sky rate agrees between the two sites, 0.755 at
Troms\o\ and 0.749 at Kiruna, while the all-hours rate does not, 0.328 and 0.481, so
the clear-sky rate is the site-independent property of the aurora and is the one on
which the amplitude of Equation~\ref{eq:global_occ} is fitted.  Above Kp~$= 2$ the training set thins:
about 50 hours per 0.5-unit bin between Kp 5 and 6 for all hours, and fewer than 20
per bin above Kp~$= 4.75$ for the clear subset, so the amplitude at high activity
rests on few observations.  Caveats on that extrapolation and
on the narrow latitude coverage are in Section~\ref{sec:limitations}.

We also tested the field's sensitivity to its driving inputs.
Perturbing the drivers of the two higher-activity hours of
Figure~\ref{fig:global_3panel} by Kp~$\pm 0.7$, $B_z \pm 3$~nT and Newell coupling
$\pm 30\%$, with the lags, rolling statistics and oval distance updated
consistently, gives a relative spread of the field of 17.4\% at Kp~$= 3.7$ and
12.1\% at Kp~$= 5.0$ over cells where the ensemble mean exceeds 0.05; at storm
level the response is mainly a shift of the ring rather than a change in amplitude.
The spans are set to the scale of ordinary variation rather than to a fitted
error: Kp~$\pm 0.7$ is two reported steps of the index, and $\pm 3$~nT is about one
standard deviation of hourly $B_z$ in the training data (2.8~nT).  The eleven
members are the central case, Kp~$\pm 0.7$, $B_z \pm 3$~nT, Newell coupling
$\pm 30\%$, and the four Kp--$B_z$ corners.
These are scenario spans rather than measured forecast errors, so this is a
sensitivity test rather than an uncertainty estimate.

\section{Discussion}
\label{sec:discussion}

\subsection{Significance of the Two-Stage Decomposition}
\label{sec:significance}

The central contribution of this work is not algorithmic (XGBoost and
logistic regression are standard methods) but rather the principled
physical decomposition of the aurora visibility prediction problem.
The cascade formula (Equation~\ref{eq:cascade}) appears straightforward
in retrospect, yet to our knowledge no prior aurora forecasting system
has explicitly implemented this decomposition.  The key requirement
is a per-hour label that distinguishes aurora identified behind cloud
cover from truly absent aurora.  The Troms\o\ AI classification system
\cite{nanjo2022automated} provides this through its AC category, which we use as a positive example of aurora
occurrence in Stage~1 training, allowing the model to learn aurora
physics with much of the weather dependence of the label removed
(Section~\ref{sec:label_s1}).

A concrete illustration of why this decomposition is necessary comes
from analyzing simultaneous observations at two of our stations.
Across the 3{,}242~overlapping hours between Kiruna and Skibotn
(2022~November--2024~February), we identified 288~hours in which
both stations recorded $y_{\mathrm{occ}}=1$ (aurora identifiable in the
images).
Of these, 107~hours (37.2\%) differed in image class between the two
sites: at one station the hour was dominated by unobscured auroral imagery
while at the other it was not.
The primary driver of this divergence is local cloud cover: 35~of
these hours showed cloud cover contrasts exceeding 70~percentage
points between the two sites, separated by only $\sim$130\,km.

\begin{figure*}[ht!]
\centering
\includegraphics[width=\columnwidth]{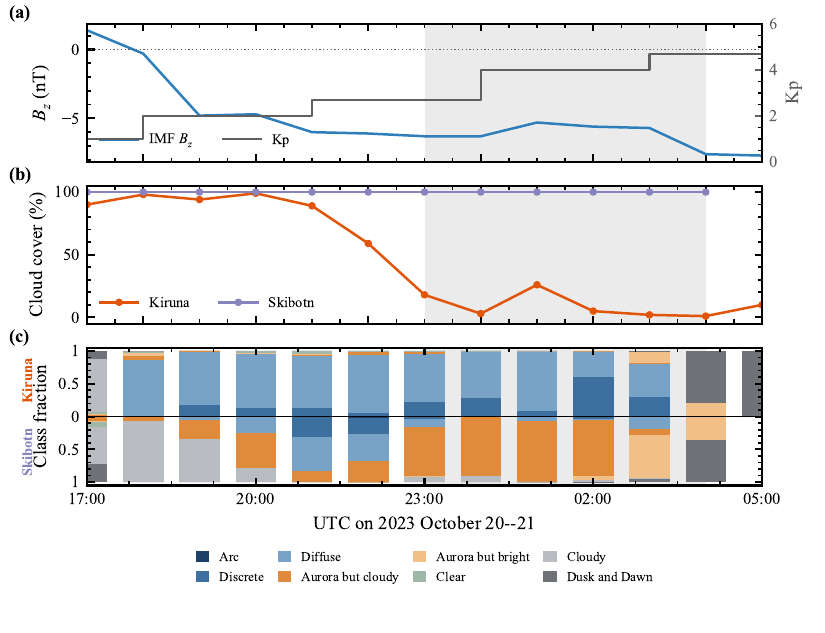}
\caption{Differing auroral image class between Kiruna and Skibotn
on 2023~October~20--21.  (a)~The drivers both sites saw: IMF $B_z$ and Kp.
(b)~Total cloud cover at each site.  (c)~Hourly image-class fractions, Kiruna above the
axis and Skibotn below, colored by the groups of Section~\ref{sec:labels}: the auroral
classes in blue, the two classes in which aurora is identified through cloud or against a
bright background in orange, and the classes holding no aurora in gray.
The gray band marks the five-hour interval
(23:00~UT October~20 to 04:00~UT October~21) during which Kiruna
was nearly clear while Skibotn remained overcast; both stations recorded
aurora throughout it.}
\label{fig:differential_case}
\end{figure*}

Figure~\ref{fig:differential_case} presents a representative event
from 2023~October~20--21.  Panel~(a) shows the shared space weather
over the displayed period; panels~(b) and~(c) show cloud cover and
aurora classification at the two sites.  The gray shaded band marks
five consecutive hours (23:00~UT October~20 to 04:00~UT October~21)
during which cloud cover diverged sharply: Kiruna was nearly
clear while Skibotn remained at 100\% overcast, even though
both stations recorded aurora activity throughout.  Such a contrast, with the same geomagnetic
forcing but opposite observing conditions at two stations only
$\sim$130\,km apart, illustrates why a single-stage visibility
model would receive contradictory supervision (the aurora recorded as
Arc, Discrete or Diffuse at the clear site and as
AC at the overcast one).  The two-stage decomposition resolves this:
Stage~1 learns aurora occurrence from space weather alone, while
Stage~2 independently predicts the divergent local observing
conditions.

The benefit of this decomposition extends beyond the accuracy
improvements in Table~\ref{tab:cascade_results}.  Stage~1 receives a
positive label whenever aurora remains identifiable in the image, including
through cloud, so hours that a visibility label would discard still contribute
to the learned physical relationships.  The recovery is partial: where
optically thick cloud hides the aurora entirely the hour is classified as
Cloudy and lost, so a period of unusually poor weather still depresses
the positive rate (Section~\ref{sec:label_s1}).

\subsection{Physical Interpretation of Feature Importance}
\label{sec:discussion_shap}

The dominance of Kp$\times$nightside (45.3\% of XGBoost importance,
13.8\% of SHAP attribution; \ref{app:shap}) reflects a
fundamental physical constraint.
At MLAT~$\approx 65$--$67^{\circ}$, the auroral oval is almost
exclusively a nightside feature during moderate activity
(Kp~$< 4$): the dayside oval sits at higher latitudes
($\sim$75$^{\circ}$ at Kp~$= 2$).  Models that use Kp alone without
the nightside indicator will systematically mis-predict aurora for
dayside hours.  The nightside indicator and the Kp$\times$nightside
interaction were supplied as features because of this known behavior; the
model's heavy reliance on them shows that the labels express the expected
dependence, not that the relation was discovered without prior
information.

The group-level SHAP analysis divides attribution across the five feature
groups as follows; these are shares of what the trained model attributes,
while predictive contribution is measured by the ablation.
Base space weather and temporal lag groups together account for
62.5\% of SHAP attribution (16.5\% and 46.0\% respectively),
providing the core geophysical predictive power; interaction terms
(13.8\%) encode the fundamental nightside and storm-phase positional
constraints; multi-station features (17.5\%) enable spatial
generalization; and derived physics contributes a modest 6.1\%,
consistent with its role as a nonlinear recombination of features
already captured by the lag group.  The leave-one-group-out ablation
(\ref{app:shap}, Table~\ref{tab:ablation}) moves validation ROC-AUC by at
most 0.016 for any group, so the model does not depend on any single
group.

\subsection{Limitations}
\label{sec:limitations}

The clearest constraint on this work is the narrow latitude range of the
available ground truth.  The three stations on which the cascade is trained
and validated lie between
MLAT 65.24$^{\circ}$ and 66.96$^{\circ}$, a span of 1.73$^{\circ}$, so cascade
performance at sub-auroral latitudes (MLAT~$< 60^{\circ}$) is
unvalidated.  Poleward of the network, the Ny-\r{A}lesund labels of
Section~\ref{sec:global_method} reach MLAT $76.8^{\circ}$ and support the
climatological level of the occurrence field there; its hour-by-hour timing that
far poleward is unvalidated.  The latitudinal structure of the hemispheric maps
(Section~\ref{sec:global_results}) comes from the Feldstein oval
parameterization rather than from behavior learned by the model.  The same
shortage of coverage applies along the activity axis: the nightside training set
holds about 50 hours per 0.5-unit Kp bin between Kp 5 and 6 and fewer than 30 above
Kp 6, so the amplitude of the hemispheric maps is extrapolated above that level, and predictions for G4--G5
storms (Kp~$\geq 8$) should be treated with caution.

The ground truth itself carries uncertainty.  Aurora occurrence labels
depend on the Troms\o\ AI ResNet-50 classifier \cite{nanjo2022automated}
(F1~$= 93.4\%$), whose imperfect accuracy propagates into our training
labels.  The AC classification is particularly challenging:
distinguishing faint aurora behind cloud from illuminated cloud is
difficult even for expert annotators.

The two stages share inputs, so their outputs are correlated.  On the Troms\o\
test set (2019--2020, $n = 6{,}144$) the Pearson correlation between the
Stage~1 and Stage~2 predicted probabilities is $r = 0.26$ ($p < 10^{-90}$),
with a Spearman rank correlation of $\rho = 0.26$.  This is expected: both
stages use MLT encodings and Stage~2 additionally uses Kp
(Section~\ref{sec:features}).  The overlap means the two stages are not
estimating fully separate quantities.

Two further limitations follow from the size and resolution of the
conditional dataset.  Stage~2 is trained on 3{,}200 aurora-occurring
hours from Troms\o\ and Kiruna (Section~\ref{sec:stage2_results}); the 1{,}587
Skibotn hours now used for evaluation would enlarge it by half, at the cost of the
withheld-site test.
The one-hour aggregation interval, in turn, averages
over substorm dynamics (10--30~minute timescales), so short-lived discrete
aurora events within a single hour are not individually resolved.

\subsection{Future Directions}
\label{sec:future}

Several extensions follow naturally from the limitations identified above.
The most immediate priority is broader latitudinal coverage, and the
oval-relative construction makes its benefit concrete.  Because latitude enters
Stage~1 only through the oval-distance coordinate, an added station does not
merely improve predictions in its own vicinity: it widens the interval of oval
distance over which the occurrence profile is constrained, and that widening
applies at every grid cell.  Resolving the profile width that
Section~\ref{sec:global_method} presently takes from the parameterization
requires sampling several oval half-widths of magnetic latitude, of order
$30^{\circ}$, whereas the current network spans $1.7^{\circ}$.  The Ny-\r{A}lesund labels of Section~\ref{sec:global_method} now supply the
poleward flank for the falloff's sign, and they quantify what remains: pinning
the width and resolving the poleward population need denser labeled records at
that latitude; the KHOnet classification of the Kjell Henriksen Observatory
record \cite{partamies2026khonet}, once released, is the natural next step.  An
equatorward site near $56$--$58^{\circ}$ (Scotland or southern Scandinavia) would
complete the span.  The binding constraint is labeled all-sky
classifications rather than the pipeline, whose other inputs are global
products.  Given accurate solar wind forecasts, the same framework produces forecasts
rather than nowcasts: pairing a solar wind model such as WSA-ENLIL
\cite{pizzo2011wsa} with numerical weather prediction cloud fields would extend it
toward multi-day planning, while the nowcast is retained as the product most
directly useful for tour operators.  WSA-ENLIL supplies solar wind speed and
density but not the interplanetary magnetic field, so the $B_z$-dependent features
would need a separate source.  The two-stage structure suits a nowcast in any case:
the observing-condition stage can be advanced on a weather forecast while the
occurrence stage is held at its current value.  On the modeling
side, recurrent architectures such as LSTMs or Transformers could capture
substorm loading--unloading dynamics more naturally than the present
hand-crafted lag features, with potential gains at sub-hourly resolution.

An operational prototype applying this two-stage framework is deployed at
\url{https://aurora-hunter.onrender.com}; \ref{app:operational} describes its
near-real-time inputs and reports a replay of the operational products against the
archival ones.

\section{Conclusions}
\label{sec:conclusions}

We have presented Aurora Hunter, a two-stage cascade that separates the
occurrence of identifiable aurora in all-sky images from the observing
conditions under which it can be seen; the cascade is evaluated retrospectively on
archival inputs.
The cascade matches a single-stage model on accuracy and is retained for what
its decomposition affords.  Cascade AUCs of
0.958 (Troms\o\ test) and 0.933 (independent Kiruna 2024) represent
improvements of $+0.095$ and $+0.097$, respectively, over the Stage~1-alone
baseline, both resolved by a DeLong test after Bonferroni correction
($p < 10^{-49}$ and $p = 1.4 \times 10^{-3}$).  This
decomposition is made possible by the ground-truth labels of the Troms\o\ AI
system: the per-hour image categories from \citeA{nanjo2022automated},
including the cloud-affected AC and bright-background AB
classes, allow Stage~1 to be trained on
image-identifiable aurora rather than on unobscured imagery alone, which
removes much of the weather dependence of the label though not all of
it.

Joint Troms\o{}+Kiruna training with 43 features (including oval distance
and seasonal encoding) achieves Stage~1 occurrence AUC~$= 0.753$ on the
independent Kiruna 2024 period and cascade visibility AUC~$= 0.933$.
At Skibotn, withheld entirely, the nightside occurrence AUC is 0.595 against
0.752 on the held-out Troms\o\ years, so transfer to a new station is
partial.
Within Stage~1, magnetic local time and the Kp$\times$nightside interaction
dominate: with the three-hour mean Kp they account for 44\% of total SHAP
attribution, encoding the constraint that aurora is a nightside phenomenon driven by
geomagnetic activity.  The normalized oval distance itself contributes 1.5\%, ranking
sixteenth.  That is expected once absolute magnetic latitude is removed from the
feature set: at a fixed station the oval distance is a function of Kp and magnetic
local time, so the model reads the same information from those directly.

For Stage~2, logistic regression and XGBoost cannot be separated on the
validation set, and we select logistic regression for its interpretable
coefficients (Section~\ref{sec:stage2_model}), which identify lunar
illumination and cloud cover as the dominant influences on the image-based
target.
Finally, the hemispheric occurrence field of Equation~\ref{eq:global_occ}
separates what is learned from what is imposed.  The activity level is learned:
it is Stage~1 on a clear-sky basis.  The oval's location, its equatorward
expansion with Kp and its day--night asymmetry are imposed by the cited
Feldstein parameterization.  One element of the spatial structure is confirmed
from our own network: the paired-station comparison of
Section~\ref{sec:global_method} establishes that occurrence falls away from the
oval center, and an independent photometric network reproduces that
sign.

Beyond the specific numerical improvements reported above, the key
conceptual contribution of this work is the principled decomposition of
aurora visibility into occurrence and observing conditions, two
processes driven by largely separate physical systems that, when modeled separately, yield
a forecast that is more interpretable and transferable and whose two
factors can be calibrated and updated independently.  This
decomposition principle is broadly applicable: any site-specific
environmental forecasting problem where a target phenomenon is
modulated by local observing conditions could benefit from a similar
two-stage architecture.

Looking forward, extending the training to stations spanning a wider
range of magnetic latitudes (Section~\ref{sec:limitations}),
incorporating numerical weather prediction cloud forecasts for
multi-day lead times, and coupling with real-time solar wind forecast
models would transform Aurora Hunter from a nowcasting tool into a
planning-horizon forecasting system for the aurora tourism industry
and space weather community alike.  The framework is an operational
proof of concept rather than a validated forecast system.  Stage~1 needs only solar wind
and geomagnetic inputs together with magnetic coordinates, and Stage~2 cloud forecasts and
ephemeris data, all of which are available operationally in near-real-time form; robust
operational use would require handling those streams, which are noisier and less complete
than the archival products evaluated here.  A prototype is deployed at
\url{https://aurora-hunter.onrender.com}
(Section~\ref{sec:future}).

\section*{Open Research}
The NASA OMNI 1-hour solar wind and geomagnetic index data used in
this study are publicly available through the Space Physics Data
Facility at \url{https://omniweb.gsfc.nasa.gov/} \cite{king2005omni}.
Hourly cloud cover reanalysis fields are available from the Open-Meteo
Historical Weather Archive API at \url{https://open-meteo.com/}
\cite{open_meteo}.  The Troms\o\ AI all-sky camera hourly classification
data are available at \url{https://tromsoe-ai.cei.uec.ac.jp/}
\cite{nanjo2022automated}.  Kiruna all-sky camera data can be obtained
from the Swedish Institute of Space Physics (IRF) at
\url{https://www.irf.se/}.  AACGM-v2 magnetic coordinate
transformations are performed using the open-source aacgmv2
Python package \cite{shepherd2014aacgm}, available at
\url{https://github.com/aburrell/aacgmv2}.
All-sky images from Ny-\r{A}lesund are from the ArcLight camera
\cite{johnsen2021arclight}, archived at NIRD \cite{arclight_nird}; the calibrated
557.7~nm intensities are those of \citeA{vandekamp2026intensities}.
The trained Aurora Hunter model weights, feature engineering code, and
evaluation scripts used in this study are openly available on Zenodo at
\url{https://doi.org/10.5281/zenodo.20446304} \cite{ge_2026_aurora_hunter_code},
together with the aligned hourly records, the human labels made for this study at
Ny-\r{A}lesund with their per-annotator returns and frame metadata, and the hourly
photometer series.

\section*{Conflict of Interest Disclosure}
The authors declare there are no conflicts of interest for this
manuscript.

\acknowledgments
This work is supported by the Shandong Provincial Natural Science Foundation
(Grant No. ZR2024MA046).
We acknowledge the University of Electro-Communications (UEC) and the
Troms\o\ Geophysical Observatory (TGO) for developing and maintaining
the Troms\o\ AI all-sky camera classification system and making
the Troms\o\ and Skibotn data publicly available
(\url{https://tromsoe-ai.cei.uec.ac.jp/};
\citeA{nanjo2022automated}).  Kiruna all-sky camera data were
provided by the Swedish Institute of Space Physics (IRF), Kiruna,
Sweden.  NASA OMNI solar wind data were obtained via the Space Physics
Data Facility (\url{https://omniweb.gsfc.nasa.gov/}).  Cloud cover
reanalysis data were obtained from the Open-Meteo Archive API
(\url{https://open-meteo.com/}).  All-sky images from
Ny-\r{A}lesund were provided by the ArcLight camera at the Light Observatory,
Ny-\r{A}lesund \cite{johnsen2021arclight}, archived at the NIRD Research Data
Archive \cite{arclight_nird}.  Calibrated 557.7~nm photometer intensities at
Kilpisj\"arvi, Abisko and Sodankyl\"a were made available by
\citeA{vandekamp2026intensities}.

Software and data analysis tools used in this study:
XGBoost \cite{chen2016xgboost}, SHAP \cite{lundberg2017shap},
Optuna \cite{akiba2019optuna}, aacgmv2 \cite{shepherd2014aacgm},
scikit-learn \cite{pedregosa2011scikit}, and Python 3.13.
Data facilities: OMNI (NASA/GSFC SPDF), Open-Meteo ERA5 reanalysis
archive, AACGM-v2, and the Troms\o\ AI all-sky camera system.

\bibliography{aurora_refs}

\appendix

\section{Feature Importance of the Two Stages}
\label{app:shap}

Feature attributions for Stage~1 are computed with SHapley additive
explanations (SHAP) \cite{lundberg2017shap} on the Troms\o\ 2018 validation set;
Table~\ref{tab:feature_importance} ranks the ten strongest features and
Table~\ref{tab:ablation} gives the feature-group ablation.

The Kp$\times$nightside interaction (45.3\% of XGBoost gain importance,
13.8\% of SHAP attribution) is the second-largest mean $|$SHAP$|$ contribution
after the MLT cosine; the bare nightside and storm indicators add almost nothing
beyond it.  It encodes the fundamental physics that aurora emission is
overwhelmingly confined to the nightside auroral oval
\cite{feldstein1967auroral,akasofu1964auroral}.  These two
interaction features, both newly introduced in this work, carry the largest
share of attribution in the trained model; their effect on predictive
performance is measured separately by the ablation below.

The two oval-distance features together contribute 3.0\% of attribution,
encoding each station's signed and absolute offset from the Kp-predicted oval
center; at a fixed station that offset is a function of Kp and magnetic local time,
which the model also reads directly.  Temporal lag features
($B_z$ lag 1h, $B_z$ min 3h) reflect the 1--3~hour substorm energy
accumulation delay, and the seasonal encoding captures the
well-known equinoctial enhancement \cite{russell1973semiannual}.

The SHAP dependences confirm the ranking: the correlation between feature value
and SHAP value is $+0.93$ for MLT$_{\cos}$ and $+0.80$ for Kp$\times$nightside,
consistent with the nightside oval preference, and $-0.76$ for $B_z$ lag 1h,
consistent with the southward IMF--substorm link.

Feature-group ablation (Table~\ref{tab:ablation}) shows that removing any
single group moves validation ROC-AUC by at most 0.016, so the model does not rest
on any one group, a desirable property for operational robustness against missing
or degraded data streams; from a single training run these margins are too small to
rank the groups against one another.

\begin{table*}
\centering
\caption{Stage~1 feature attribution, ordered by mean $|$SHAP$|$ on the
  Troms\o\ 2018 validation set, with the rank the same feature takes under
  XGBoost gain importance.  Features marked $\star$ are newly introduced in this
  work; letters give the feature group of Table~\ref{tab:features}.
  \label{tab:feature_importance}}
{\small
\begin{tabularx}{\textwidth}{rlrrcX}
\toprule
Rank & Feature & mean$|$SHAP$|$ & \% total & XGB rank & Physical interpretation \\
\midrule
 1 & MLT$_{\cos}$                 & 0.815 & 22.6\% &  1 &
   Cyclic encoding of magnetic local time \\
 2 & Kp$\times$nightside$^\star$  & 0.497 & 13.8\% &  2 &
   Geomagnetic activity weighted by nightside position \\
 3 & Kp mean 3h                   & 0.281 &  7.8\% &  3 &
   Smoothed geomagnetic activity \\
 4 & Kp                           & 0.236 &  6.5\% &  4 &
   Planetary geomagnetic activity index \\
 5 & season\_sin$^\star$          & 0.160 &  4.4\% &  5 &
   Seasonal cycle encoding \\
 6 & $B_z$ lag 1h                 & 0.113 &  3.1\% &  6 &
   Southward IMF one hour prior \\
 7 & equinox\_sin$^\star$         & 0.109 &  3.0\% &  7 &
   Russell--McPherron semiannual encoding \\
 8 & season\_cos$^\star$          & 0.102 &  2.8\% &  8 &
   Seasonal cycle encoding \\
 9 & Newell sum 4h                & 0.092 &  2.6\% &  9 &
   Four-hour accumulated coupling function \\
10 & Kp mean 6h                   & 0.087 &  2.4\% & 10 &
   Longer-term activity level \\
\bottomrule
\end{tabularx}
}
\end{table*}

The strongest SHAP interaction pairs all involve the MLT cosine: with
Kp$\times$nightside (0.182), Kp (0.103), the seasonal cosine (0.099) and the
three-hour mean Kp (0.096), followed by Kp$\times$nightside with the oval
distance (0.072); activity and position are read jointly.

\begin{table*}
\centering
\caption{Stage~1 feature-group ablation, in two complementary views.  Left:
  leave-one-group-out, each row removing the named group from the full
  43-feature model.  Right: cumulative addition, each row adding the named
  group to those above it.  $\Delta$ is the change in validation ROC-AUC
  against the reference row of that view.
  \label{tab:ablation}}
{\footnotesize\setlength{\tabcolsep}{4pt}
\begin{tabular}{lrrrr@{\hskip 1em}lrrrr}
\toprule
\multicolumn{5}{c}{Leave one group out} & \multicolumn{5}{c}{Cumulative addition} \\
\cmidrule(r{1.5em}){1-5}\cmidrule{6-10}
Configuration & $N$ & Val & $\Delta$ & Test &
Configuration & $N$ & Val & $\Delta$ & Test \\
\midrule
Full model (all 43)     & 43 & 0.856 &          & 0.849 &
(a) Base SW+coords & 10 & 0.864 &          & 0.831 \\
w/o base SW+coords & 33 & 0.852 & $-0.004$ & 0.848 &
+ (b) Temporal lags     & 24 & 0.873 & $+0.009$ & 0.844 \\
w/o temporal lags       & 29 & 0.846 & $-0.010$ & 0.847 &
+ (c) Derived physics   & 32 & 0.876 & $+0.003$ & 0.843 \\
w/o derived physics     & 35 & 0.854 & $-0.002$ & 0.849 &
+ (d) Interaction       & 35 & 0.872 & $-0.004$ & 0.843 \\
w/o interaction         & 40 & 0.853 & $-0.003$ & 0.833 &
+ (e) Multi-station     & 43 & 0.856 & $-0.016$ & 0.849 \\
w/o multi-station       & 35 & 0.872 & $+0.016$ & 0.843 & & & & & \\
\bottomrule
\end{tabular}
}
\end{table*}

\begin{figure}[ht!]
\centering
\includegraphics[width=\columnwidth]{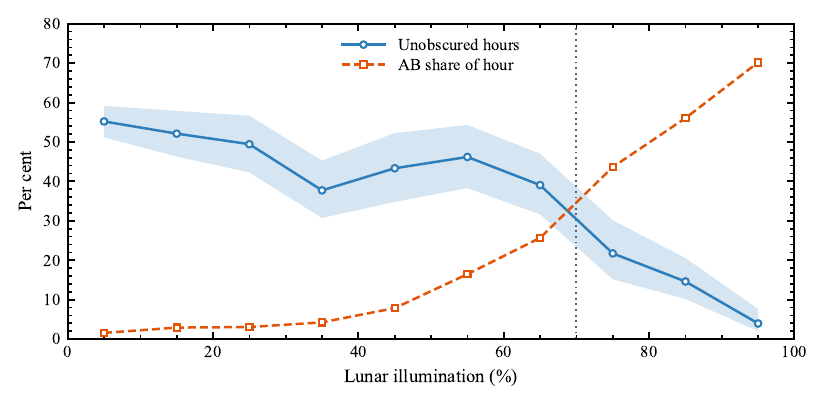}
\caption{Effect of lunar illumination on the unobscured-classification
  rate (Troms\o{}, aurora-occurring hours 2015--2020).  Points give
  the rate in each 10\% illumination bin and the band the Wilson 95\% interval; bins hold
  between 115 and 594 of 2{,}136 hours.  The dotted vertical line marks the
  70\% threshold used as a binary feature in Stage~2.  The rate falls from 55.2\% at
  illumination below 10\% to 21.7\% in the 70--80\% bin and 8.9\% above 80\%, motivating
  the nonlinear threshold feature.  The second series is the mean
  AB share of the hour, which rises from 1.6\% to 63.6\% across the same range
  and crosses the first at the threshold: what falls with illumination is the probability
  of a clean classification, not the ability to detect aurora.  Between 30\% and 70\%
  illumination the intervals overlap, so the variation across those bins is not
  resolved.}
\label{fig:moon_effect}
\end{figure}

For Stage~2, the standardized LR coefficients (Table~\ref{tab:stage2_coefs})
have direct physical interpretations.
The moonlight term $B_{\rm sky}$ (coefficient $-1.99$) is the strongest
single predictor of an obscured
classification, ahead of mid-level cloud ($-0.90$): at illumination
$>80\%$, the unobscured-classification rate drops to only 8.9\%
(Figure~\ref{fig:moon_effect}).  This decline is in part a property of the
label: aurora recorded against a moonlit sky is classified as AB and
counts as a Stage~2 negative, so what falls with illumination is the probability
of an unobscured classification rather than the ability to detect aurora.  This severe
nonlinear threshold, not
captured by raw illumination percentage alone, justifies the binary
high-illumination flag as a feature.
Mid-altitude cloud cover (coefficient $-0.900$) has a larger coefficient
magnitude than low-altitude cloud cover ($-0.544$), likely because
mid-altitude cloud is more spatially uniform and optically thick along the line of
sight to the emission layer.  Post-midnight position ($+0.371$)
is the strongest positive MLT predictor, reflecting the well-established
prevalence of bright, structured discrete aurora in the post-midnight
sector \cite{akasofu1964auroral}.

\begin{table*}
\centering
\caption{Stage~2 LR coefficients (standardized, top 10 by $|$coef$|$).
  Positive values improve the chance of an unobscured classification; negative values reduce it.
  \label{tab:stage2_coefs}}
\small
\begin{tabular}{rlrl}
\toprule
Rank & Feature & Coeff. & Physical effect \\
\midrule
 1 & sky\_brightness   & $-1.986$ & Moonlight illuminating the sky, zero below the horizon \\
 2 & cloud\_cover\_mid & $-0.900$ & Mid-altitude cloud obscuration \\
 3 & cloud\_cover\_low & $-0.544$ & Low cloud obscuration \\
 4 & is\_postmidnight  & $+0.371$ & Diffuse/discrete aurora sector \\
 5 & cloud\_cover\_high & $-0.330$ & High-altitude cirrus \\
 6 & high\_illum       & $-0.284$ & Bright background shifts images to AB \\
 7 & cloud\_opacity    & $+0.262$ & Opacity-weighted cloud sum, collinear with the three layer terms \\
 8 & MLT$_{\cos}$       & $+0.219$ & Midnight-centered preference \\
 9 & Kp\_intensity     & $+0.214$ & Brighter aurora through thin cloud \\
10 & is\_premidnight   & $+0.193$ & Discrete aurora sector \\
\bottomrule
\end{tabular}
\end{table*}

\section{Per-Year Cascade Performance}
\label{app:yearly}

\begin{figure*}[ht!]
\centering
\includegraphics[width=\columnwidth]{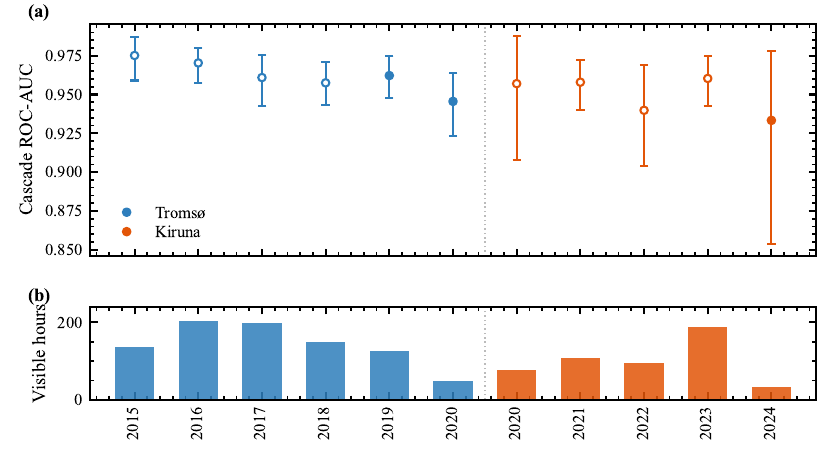}
\caption{(a)~Per-year cascade ROC-AUC with bootstrap 95\% confidence
  intervals (24-hour blocks, 2000 resamples).  Open symbols mark years used in
  training or validation, filled symbols years held out from both stages.
  (b)~Aurora-visible hours in each year, the count the interval above each one
  rests on.
  \label{fig:yearly_auc}}
\end{figure*}

Figure~\ref{fig:yearly_auc} reports cascade ROC-AUC year by year for
Troms\o\ (2015--2020) and Kiruna (2020--2024), with the number of
aurora-visible hours in each year and a bootstrap 95\% confidence interval
obtained by resampling 24-hour blocks.  Three years are held out from both
stages: Troms\o\ 2019 and 2020, which form the test set, and Kiruna 2024, the
independent period at the second site.  They give 0.962, 0.946 and 0.933, each
interval covering their pooled value of 0.954, and the training and validation
years fall between 0.940 and 0.975.  Across all eleven years every interval
overlaps every other, so performance is consistent over the record.  Annual
event counts run from 32 to 203, and the intervals are correspondingly wide:
Kiruna 2024 rests on 32 aurora-visible hours and spans 0.854 to 0.978, so the
figure shows consistency over the record rather than resolving differences
between years.

\section{The Operational Prototype and an Input Replay}
\label{app:operational}

A prototype applying this two-stage framework is deployed at
\url{https://aurora-hunter.onrender.com}.  It ingests near-real-time inputs rather than the
archival products used throughout this paper: solar wind and interplanetary magnetic field
from the NOAA Space Weather Prediction Center real-time solar wind feed at one-minute
cadence, the Center's estimated Kp at its three-hour cadence, real-time
Dst, and numerical weather prediction cloud forecasts over the 12-hour nowcast horizon, with lunar geometry
computed rather than retrieved.  Locations outside the magnetic latitude band spanned by
the training stations are flagged as extrapolation and the occurrence stage is not applied
there.  The prototype now runs the model documented in this paper: the 43-feature occurrence
stage with its Platt calibration, the clear-sky amplitude and Feldstein profile of
Section~\ref{sec:global_method}, and the 15-feature observing-conditions stage.  The prototype departs
from that model in one respect, deliberately: the lowest block of the Stage~2 isotonic calibrator
returns exactly zero, and the deployed version floors it at the rate observed in those
hours, on the grounds that a public tool should not report a clean view as impossible.

Because those inputs differ from the final products this paper evaluates, the skill
reported in Section~\ref{sec:results} can be regarded as an upper bound on what the
prototype achieves.  How far the two chains diverge can be measured for
the space weather half of the inputs, by reconstructing the snapshot a live system would
have held and running Stage~1 on it.  We did so for Skibotn, the station withheld from
training entirely, over 2025 January to April, using the DSCOVR real-time magnetometer and
Faraday cup, the published three-hour Kp, and provisional Dst in place of the definitive
OMNI series.

\begin{figure}[ht!]
\centering
\includegraphics[width=\columnwidth]{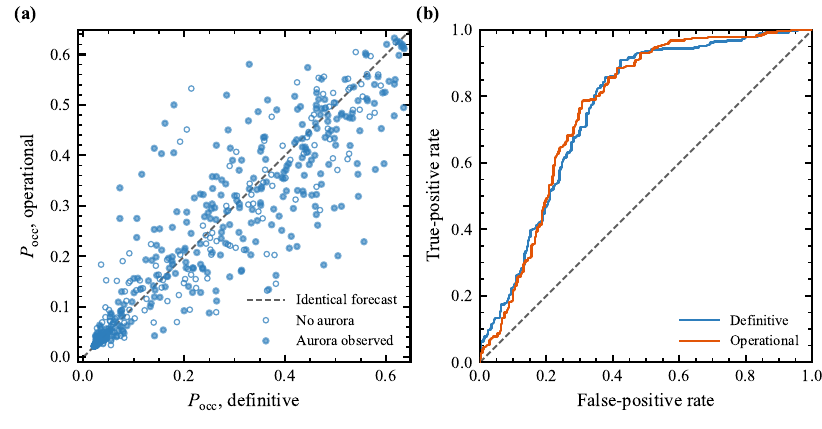}
\caption{Stage~1 occurrence probability at Skibotn over the same 571 hours,
  computed from the archival inputs used throughout this paper and from the products
  a live system would have held.  \textit{(a)}~The two agree hour by hour; the dashed
  line is equality, open symbols are hours scored aurora-free and filled symbols
  hours with aurora.  \textit{(b)}~ROC on the same hours.  The quantity this figure
  establishes is the paired difference between the two chains, given in the text;
  the two absolute values are computed on a selected subset and are not comparable
  with the performance reported elsewhere in the paper.
  \label{fig:operational_replay}}
\end{figure}

The two chains agree closely (Figure~\ref{fig:operational_replay}).  Hour by hour the
occurrence probabilities correlate at $r = 0.91$, and on the same hours the operational
chain reaches ROC-AUC $0.767$ against $0.763$ for the archival chain, a paired difference of
$+0.004$ with a 95\% bootstrap interval of $[-0.011, +0.020]$.  Real-time solar wind and
index products therefore incur essentially no loss of ranking skill relative to the
definitive series.  Two limits on that statement should be read with it.  It concerns
Stage~1 only, so it says nothing about replacing the cloud reanalysis with a forecast, which
is the larger of the two gaps and remains unquantified.  And it is measured on hours in which the
camera could see the sky, which is a different population from the one Section~\ref{sec:results}
evaluates.

\end{document}